\documentclass[%
 reprint,
 superscriptaddress,
 amsmath, 
 amssymb,
 amsthm,
 aps,
 pra,
]{revtex4-2}
\usepackage{graphicx}
\usepackage{dcolumn}
\usepackage{bm}

\usepackage[english]{babel}
\makeatletter
\let\ORIbbl@fixname\bbl@fixname
\def\bbl@fixname#1{%
  \@ifundefined{languagealias@\expandafter\string#1}
    {\ORIbbl@fixname#1}
    {\edef\languagename{\@nameuse{languagealias@#1}}}%
}
\newcommand{\definelanguagealias}[2]{%
  \@namedef{languagealias@#1}{#2}%
}
\makeatother
\definelanguagealias{en}{english}
\definelanguagealias{eng}{english}

\usepackage{enumitem} 
\usepackage{braket} 
\usepackage{mathtools} 
\usepackage[dvipsnames]{xcolor}
\usepackage{comment} 
\usepackage{physics} 

\usepackage[font=small, justification=Justified]{caption}
\usepackage{subcaption}

\newcommand{\eqeq}{\overset{\mathrm{eq}}{=}}

\newcommand{\mbeq}{\overset{!}{=}}

\usepackage{amsthm}
\theoremstyle{definition}
\newtheorem{definition}{Definition}[section]

\definecolor{S_Blue}{RGB}{0,135,252}
\definecolor{S_Red}{RGB}{214,13,63}
\definecolor{Blue}{RGB}{47,89,151}
\definecolor{S_Grey}{RGB}{150,150,158}
\definecolor{S_Yel}{RGB}{255,204,0}
\definecolor{S_Green}{RGB}{102,204,0}
\definecolor{S_Brown}{RGB}{154,41,41}

\usepackage{mathrsfs} 

\usepackage{stmaryrd} 

\usepackage{booktabs} 

\usepackage{pgffor}
\usepackage{ifthen}
\usepackage{placeins}
\usepackage[unicode=true]{hyperref} 

\begin{document}

\preprint{APS/123-QED}

\title{Observable Statistical Mechanics}

\author{Lodovico Scarpa}
 \email{lodovico.scarpa@physics.ox.ac.uk}
 \affiliation{%
 Clarendon Laboratory, University of Oxford, Parks Road, Oxford OX1 3PU, UK
 }%

 \author{Abdulla Alhajri}
  \affiliation{%
 Clarendon Laboratory, University of Oxford, Parks Road, Oxford OX1 3PU, UK
 }%
 \affiliation{%
 Quantum Research Centre, Technology Innovation Institute, 9639 Abu Dhabi, United Arab Emirates
 }%
 
\author{Vlatko Vedral}%
 \affiliation{%
 Clarendon Laboratory, University of Oxford, Parks Road, Oxford OX1 3PU, UK
 }
 
 \author{Fabio Anza}
 \email{fanza@umbc.edu}
  \affiliation{
 Department of Physics, Cybersecurity Institute, Quantum Science Institute, University of Maryland, Baltimore County, Baltimore, Maryland, US
 }

\date{Jul 29, 2025}

\begin{abstract}
\noindent Predicting the stationary behavior of observables in isolated many-body quantum systems is a central challenge in quantum statistical mechanics. While one can often use the Gibbs ensemble, which is simple to compute, there are many scenarios where this is not possible and one must instead use another ensemble, such as the diagonal, microcanonical or generalized Gibbs ensembles. However, these all require detailed information about the energy or other conserved quantities to be constructed. Here we propose a general and computationally easy approach to determine the stationary probability distribution of observables with few outcomes. Interpreting coarse measurements at equilibrium as noisy communication channels, we provide general analytical arguments in favor of the applicability of a maximum entropy principle for this class of observables. We show that the resulting theory accurately predicts stationary probability distributions without detailed microscopic information like the energy eigenstates. Extensive numerical experiments on 7 non-weakly interacting spin-1/2 Hamiltonians demonstrate the broad applicability and robustness of this framework in both quantum integrable and chaotic models.
\end{abstract}

\maketitle

Understanding and predicting the stationary behavior of isolated many-body quantum systems is one of the central challenges in quantum physics. Prima facie, it does not seem possible for these systems to reach equilibrium, as they evolve under fully unitary dynamics, which are reversible, quasi-periodic, distance-preserving and leave the von Neumann entropy invariant, among other features. Nevertheless, it is well known that under generic conditions observables will exhibit effective equilibration, wherein their expectation values keep on evolving yet remain very close to their time average for the vast majority of times \cite{reimann_foundation_2008, linden_quantum_2009, reimann_equilibration_2012, binder_equilibration_2018, gogolin_equilibration_2014, eisert_quantum_2015, garcia-pintos_equilibration_2017}. When this happens, the most accurate description of their stationary behavior is given by the totally dephased state in the energy basis (the ``diagonal ensemble") \cite{binder_equilibration_2018, gogolin_equilibration_2014}. However, constructing this state requires knowledge of the energy eigenvectors, and thus quickly becomes computationally prohibitive due to the exponential growth of the Hilbert space. Often observables do not merely reach a steady state but actually thermalize, which is a stronger requirement that also implies independence from the initial conditions. When this happens, their stationary value can be predicted using standard statistical ensembles. In a narrow microcanonical window around a given energy, the suitable ensemble is the microcanonical one, which however also demands knowledge of an exponential number of energy eigenstates to be computed. For local observables in large and weakly interacting systems with exponential density of states, predictions at thermal equilibrium can be greatly simplified by employing the Gibbs ensemble, which only requires local information. Nevertheless, there exist many physically relevant scenarios for which this paradigm does not apply. Examples include integrable systems, described by generalized Gibbs ensembles \cite{essler_quench_2016, vidmar_generalized_2016}; strongly interacting systems, characterized by the mean-force Gibbs state \cite{miller_hamiltonian_2018}; non-local observables; as well as phenomena such as pre-thermalization \cite{mori_thermalization_2018}, quantum scars \cite{serbyn_quantum_2021, moudgalya_quantum_2022}, Hilbert space fragmentation \cite{moudgalya_quantum_2022}, Many-Body Localization \cite{nandkishore_many-body_2015, abanin_recent_2017, alet_many-body_2018} and dynamical symmetries \cite{buca_non-stationary_2019}. These cases highlight the intricate dynamics at play, and thus the need for a more unified and nuanced approach. In particular, we argue this approach should not just focus on density matrices, but instead take into consideration the interplay of all three actors, namely, the observable, the initial state and the Hamiltonian. This allows to fully characterize the problem mathematically, while also bringing the theory closer to experimental reality, where we typically have access to only a few observables and cannot probe the entire density matrix. \\

In this paper, we introduce Observable Statistical Mechanics, a framework that allows to easily predict the stationary behavior of a large class of physically accessible observables, namely, those with a limited number of outcomes (so-called ``coarse" or ``highly degenerate" observables). The theory relies on the constrained maximization of the entropy of an observable's probability distribution.  We provide analytical arguments for the validity of this maximum entropy principle, and numerically show that it enables accurate predictions of stationary measurement statistics without strong assumptions on the Hamiltonian or knowledge of the underlying density matrix, in both chaotic and integrable systems. Crucially, it does so without detailed knowledge of energy eigenstates, or any extensive set of conserved quantities for integrable systems. In addition, our predictions work even beyond weak coupling conditions and far away from the thermodynamic limit, thus going beyond the regime of applicability of equilibrium statistical mechanics. \\
We stress that here we are not attempting to understand the necessary and sufficient conditions for the emergence of thermal equilibrium, and hence our work represents a significant departure from the existing literature, such as in quantum typicality \cite{lloyd_pure_2013, tasaki_quantum_1998, hayden_aspects_2006, popescu_entanglement_2006, goldstein_canonical_2006, goldstein_approach_2010, goldstein_long-time_2010, masanes_complexity_2013, malabarba_quantum_2014, goldstein_extremely_2015, reimann_typical_2016, binder_dynamical_2018, binder_equilibration_2018}, Eigenstate Thermalization Hypothesis \cite{dalessio_quantum_2016, deutsch_eigenstate_2018, srednicki_chaos_1994, srednicki_approach_1999, rigol_thermalization_2008} and Random Matrix Theory \cite{guhr_random-matrix_1998, borgonovi_quantum_2016, khaymovich_eigenstate_2019, weidenmuller_random-matrix_2024}. Rather, we are offering a general and computationally easy approach to estimate stationary measurement statistics for a large class of experimentally accessible observables. Nevertheless, we expect that our results could inform the foundations of statistical mechanics, particularly our analytical results on the informational capacity and equilibrium entropy of highly degenerate observables. \\

\paragraph*{Related literature} Among many relevant works, here we mention three related research directions. Firstly, the focus on \emph{Observational Entropy} \cite{safranek_quantum_2019, strasberg_first_2021, safranek_brief_2021}. This is a generalization of observable entropy as the Shannon entropy of a generic Hilbert space partition. In subsection \ref{subsec:EEs_derivation} we actually justify physically the emergence of this thermodynamic entropy as the equilibrium entropy for highly degenerate few-body observables. Moreover, after the first version of this manuscript, work done in Ref.~\cite{meier_emergence_2024} has found additional evidence for the dynamical maximization of observable entropy. While limited to one Hamiltonian model in their numerical simulations, that work supports the new theoretical picture we develop here. \\
Secondly, the field of quantum typicality. The core idea of this approach is to consider mathematically typical conditions, i.e., to choose a Haar random initial state, Hamiltonian or observable. This then allows to prove statements applying to the overwhelming majority of cases via concentration of measure arguments \cite{ledoux_concentration_2020, popescu_entanglement_2006}. In particular, using these methods it has been shown that generically equilibration and thermalization are typical phenomena, and that the typical relaxation timescales are extremely short \cite{lloyd_pure_2013, tasaki_quantum_1998, hayden_aspects_2006, popescu_entanglement_2006, goldstein_canonical_2006, goldstein_approach_2010, goldstein_long-time_2010, masanes_complexity_2013, malabarba_quantum_2014, goldstein_extremely_2015, reimann_typical_2016, binder_dynamical_2018, binder_equilibration_2018}. Despite their mathematical rigor, these arguments can often lead to physically absurd conclusions, such as that a cup of coffee cools down in less than a microsecond \cite{goldstein_extremely_2015}. This means that in many cases the physical situations of interest correspond to mathematically \textit{un}typical ones. Hence, while these results suggest that under typical conditions the entropy of coarse observables is maximized and small subsystems are well described by the Gibbs ensemble, we stress that here we are not assuming any randomness. Moreover, we remind the reader about the several aforementioned physical scenarios where it is known that the Gibbs ensemble cannot be used. \\
Finally, we mention a third research direction, focused on deriving effective equations of motion to understand relaxation to a stationary state. Important examples are the Quantum Generalized Hydrodynamics \cite{ruggiero_quantum_2020} and the more recent work by Strasberg and collaborators \cite{strasberg_classicality_2023} aimed at deriving an effective master equation for observables which are both \emph{slow} and \emph{coarse}. In Section \ref{sec:discussion} we make an explicit connection with Ref. \cite{strasberg_classicality_2023}.

\section{Results}\label{sec:results}
\subsection{Setup} \label{subsec:setup}
We consider a many-body quantum system made by $N$ interacting units, each of dimension $d$, and thus with Hilbert space $\mathcal{H}$ of dimension $D = d^N$. The entire system evolves with Hamiltonian $H=\sum_n E_n \ketbra{E_n}$, and it is initialized in a pure state $\ket{\psi_0} = \sum_n c_n \ket{E_n} \in \mathcal{H}$. Our system can be probed via a series of compatible projective measurements corresponding to an observable $A := \sum_{j=1}^{n_A} a_j A_j$ with eigenprojectors $A_j := \sum_{s=1}^{d_j} |j, s \rangle\langle j,s|$, where $\left\{\ket{j,s}\right\}_{j=1,s=1}^{n_{A},d_j}$ is a generic basis diagonalizing our observable $A$. This identifies a decomposition of the full Hilbert space into $n_A$ subspaces $\left\{\mathcal{H}_j\right\}_{j=1}^{n_A}$, each of dimension $d_j := \mathrm{dim} \mathcal{H}_j = \Tr A_j$, in which the measurement has a definite outcome. While here we focus on projective measurements, the whole framework can be extended straightforwardly to generic measurements described by a Positive Operator-Valued Measure (POVM). Moreover, hereafter we will focus on highly degenerate observables. Indeed, in realistic experiments, the number of classically distinguishable outcomes is limited compared to the full dimension of the Hilbert space $n_{A} \ll D$. Since $\sum_{j=1}^{n_{A}}d_j = D$, this means we are always in the situation in which $\log d_j \gg 1$ and, in most cases, $\log d_j \sim N$. Typical examples are von Neumann macro-observables and observables with support on less than half of the entire system \cite{anza_eigenstate_2018}.

The time-dependent probability distribution of the measurement outcomes $p_j(t)\coloneqq \bra{\psi_t}A_j\ket{\psi_t}$ evolves as a result of the global unitary dynamics as
\begin{equation}
    p_j(t):=\sum_{n,m} c_n^{*} c_m e^{\frac{i}{\hbar}(E_n-E_m)t} \left[ A_j\right]_{nm},
\end{equation}
with $[A_j]_{nm}:=\bra{E_n}A_j\ket{E_m}$. When the system is many-body and interacting, the energy spectrum is usually non-degenerate and has non-degenerate energy gaps \cite{goldstein_long-time_2010, von_neumann_proof_2010, linden_quantum_2009, srednicki_approach_1999, reimann_foundation_2008, short_quantum_2012, reimann_equilibration_2012}. These conditions lead to a central result in the study of stationarity in closed quantum systems \cite{gogolin_equilibration_2014}: \emph{If a system equilibrates, it will do so to the predictions of the diagonal ensemble} $\rho_{DE}:=\sum_n |c_n|^2 \ketbra{E_n}$. We stress that when we talk about ``equilibration" we really just mean the phenomenon of reaching a stationary state, rather than the stronger notion of thermal equilibrium. Defining $p_j^{DE}:= \Tr \left(\rho_{DE}A_j\right) = \sum_n |c_n|^2 [A_j]_{nn}$, we indicate this phenomenology with the following notation
\begin{equation}
    p_j(t) \rightsquigarrow p_j^{DE}~.
\end{equation}
At the technical level, both the non-degenerate energy spectrum and non-degenerate energy gaps assumptions can be relaxed \cite{short_quantum_2012, reimann_equilibration_2012, masanes_complexity_2013, balz_equilibration_2016}. Direct computation of $\rho_{DE}$ is often unfeasible as it requires knowledge of an exponentially large number of energy eigenstates. The issue of understanding and predicting the stationary distributions then becomes: \emph{How can we estimate $p_j^{DE}$ without the energy eigenstates?} As discussed in the introduction, although the Gibbs ensemble can be used in some cases, there are other scenarios where it is not applicable. Observable Statistical Mechanics provides a general and principled answer, with abundant analytical and numerical evidence supporting it---the \emph{Maximum Observable Entropy Principle} (MOEP). Beyond the notation that has already been introduced, the main actors are the following. $S(H)=-\sum_{n}|c_n|^2 \log |c_n|^2 $ is the entropy of the energy distribution, also called the Diagonal Entropy \cite{polkovnikov_microscopic_2011}.  $S(A):= - \sum_{j=1}^{n_A} p_j^{DE} \log p_j^{DE}$ is the entropy of the stationary measurement statistics of $A$ and $\tilde{S}(A) = -\sum_{j=1}^{n_A} \sum_{s=1}^{d_j} p_{j,s}^{DE}\log p_{j,s}^{DE}$ is the entropy of $p_{j,s}^{DE}:=\bra{j,s}\rho_{DE}\ket{j,s}$, where $\left\{\ket{j,s}\right\}$ is a generic basis diagonalizing $A$. The two are related via $p_j^{DE} = \sum_{s=1}^{d_j}p_{js}^{DE}$. Additionally, $S(H,A):= - \sum_{j=1}^{n_A}\sum_{n=1}^{d^N}p_{j,n}^{DE} \log p_{j,n}^{DE}$ is the entropy of the joint distribution $p_{j,n}^{DE}$ obtained by measuring first the energy $H$ and then $A$ on the stationary state.

\subsection{Maximum Observable Entropy Principle} \label{subsec:MOEP}
The issue of estimating $p_j^{DE}$ while lacking complete information (the energy eigenstates) can be phrased as a standard inference problem. While most efforts in the literature have been focused on estimating or approximating the entire $\rho_{DE}$ (e.g. the GGE construction \cite{essler_quench_2016, vidmar_generalized_2016}), this approach is often too demanding and can easily fail: it is sufficient that one observable does not agree with the prediction. It also does not consider that, in most experiments, we only have access to a limited number of observables. Instead, inferring directly the probability distribution of measurement outcomes is a minimal and more practical route. This leads to the following principle.
\begin{definition}[Maximum Observable Entropy Principle]
\emph{The observed stationary measurement statistics $p_j^{DE}$ is well described by a distribution obtained by maximizing the entropy $\tilde{S}(A)$ under perturbations with conserved average energy.}
\end{definition}
While first formulated to study the emergence of thermal equilibrium \cite{anza_information-theoretic_2017, anza_eigenstate_2018}, as we show explicitly in the following, its predictive capabilities go well beyond its initial intent. We begin by discussing a simple heuristic argument that points towards the validity of MOEP in highly degenerate observables. This will help the reader gain intuition on this principle. We then present analytical predictions of MOEP and showcase their validity in a large-scale numerical experiment.

\subsection{Validity of MOEP} \label{subsec:heuristics}
The posited Maximum Observable Entropy Principle states that our best guess for $p_j^{DE} = \sum_s p_{js}^{DE}$ is the distribution resulting from the constrained maximization of $\tilde{S}(A)$. In principle, all linearly independent conserved quantities should be included as constraints ---normalization and the expectations of all energy eigenstates $|c_n|^2$. Including all of them invariably leads to the exact result $p_j^{DE}$ \cite{anza_new_2018}, but it requires knowledge of all energy eigenstates. The reason for the emergent validity of MOEP is that we do not need to include all the $|c_n|^2$. As we argue now, this occurs for two reasons: narrowness of energy distribution and low accessible information.

\subsubsection{Narrowness of the energy distribution}
As convincingly argued by Reimann \cite{reimann_foundation_2008,reimann_equilibration_2012-1}, realistic and thermodynamically stable initial states of a mesoscopic system have an energy distribution that is localized within a microcanonical window: $|c_n|^2 \neq 0$ only for $E_n \in \mathcal{I}_{mc}(E,\Delta E):=[E - \frac{\Delta E}{2},E + \frac{\Delta E}{2}]$. The window is macroscopically small $\Delta E \ll |E_{max}-E_{min}|$ but it still contains an exponentially large number of energy eigenstates. Inside $\mathcal{I}_{mc}$, therefore, it will be incredibly difficult for experimentalists to gain control over single energy eigenstates with microscopically small energy differences. We therefore expect the entropy of the energy distribution to scale linearly with system size $S(H) \sim k_E N$ but with a relatively small coefficient: $S(H) \ll N \log d$: $k_e \ll \log d$. This is a request of energetic stability of the initial states that we can meaningfully prepare. But this alone is by no means sufficient to prove the emergence of standard thermal equilibrium states or the validity of statistical mechanics.

\subsubsection{Low accessible information} 
Leveraging information and communication theory \cite{nielsen_quantum_2010}, the operation of measuring the observable $A$ on the stationary state can be understood as a communication channel in which the source is sending orthogonal energy eigenstates $\ket{E_n}$, each with probability $|c_n|^2$, through a measurement channel determined by the spectral decomposition of $A$. The amount of information accessible after the measurement is given by $I(H,A) := S(H) + S(A) - S(H,A) \leq S(H)$ where the rightmost inequality is Holevo's bound \cite{nielsen_quantum_2010}. The measurement is very informative when Holevo's bound is saturated, so that the receiver has access to the maximal amount of information sent by the source: $I(H,A) \approx S(H)$. Otherwise, when $I(H,A) \ll S(H)$ the channel is very noisy and the measurement not informative. Most observables of interest $A$ correspond to measurements of the latter kind, carrying very little details about the stationary state. Indeed, while the source can emit up to $2^{S(H)} \leq d^N$ orthogonal states, realistic measurements yield a small amount of information about the source: $2^{S(A)} \leq n_A \ll d^N$. The following inequality helps us quantify this strong asymmetry between the amount of information produced by the source and the one available to the receiver: $\frac{I(H,A)}{S(H)} \leq \frac{S(A)}{S(H)} \leq \frac{\log n_A}{S(H)} \ll 1$. The rightmost inequality holds in a large number of experimentally realistic situations, as the following two examples clarify. First, measurements with support on few-bodies $N_S \ll N$ yield $\frac{\log n_A}{S(H)} \sim \frac{N_S}{N} \ll 1$. Second, the result is not limited to measurements on few-body operators, although the scaling is more favourable there. Global operators resulting from uniform lattice averages of $k$-body operators, like the total magnetization in a spin-$1/2$ system or the total number of electrons in a lattice, have at most a number of eigenvalues scaling linearly with system size: $n_A \in \mathcal{O}(N)$, leading again to $\frac{\log n_A}{S(H)} \sim \frac{\log N}{N} \ll 1$. 
This proves that in most experimentally reasonable measurements we stand in a situation of \emph{low accessible information}: $I(H,A) \ll S(H)$.

\subsubsection{Extensivity of the entropy} 
We established that the stationary measurement statistics $p_j^{DE}$ contains very little information about the microscopic details of $\rho_{DE}$. Adapting Jaynes' original max entropy principle \cite{jaynes_information_1957, jaynes_brandeis_1989} to measurement statistics rather than the entire state of the system leads to the Maximum Observable Entropy Principle. In other words: a minimal amount of knowledge of the stationary state is sufficient to estimate $p_j^{DE}$. Owing to the concentration of the energy distribution $|c_n|^2$, in most cases of practical relevance, the average energy $E = \sum_{n}|c_n|^2 E_n$ should be enough. While these information-theoretic arguments fully justify the use of MOEP, here we strengthen them by showing the entropy $\tilde{S}(A)$ is extensive. Given a basis that diagonalizes our observable $\left\{\ket{j,s}\right\}_{j=1,s=1}^{n_{A},d_j}$, by noting that $p_{js}^{DE}$ are the diagonal elements of $\rho_{DE}$ in the eigenbasis $\{|j,s\rangle\}_{j,s}$ and by using the Schur-Horn theorem, we have that $\{p_{js}\}$ majorizes $\{|c_n|^2\}$. Due to the Schur-concavity of the Shannon entropy, at equilibrium it always holds that $\tilde{S}(A) \geq S(H)$. Hence, if $S(H) \sim k_E N$, then $\tilde{S}(A)$ is also extensive. As detailed in the Methods (subsection \ref{subsubsec:extensive_entropy}), we can make this statement stronger for few-body observables by showing that in this case $\tilde{S}(A) \geq (N - N_S) \log d$, where $N_S$ is the size of the support of the observable.
Such linear scaling of the observable entropy is the hallmark of an \emph{observable statistical mechanics}: a statistical mechanics of measurement outcomes rather than states. 
This concludes our discussion in support of MOEP. We now move on to explain the ensuing theory: Observable Statistical Mechanics.

\subsection{Observable Statistical Mechanics} \label{subsec:results}

We present the complete theory of Observable Statistical Mechanics, with its main predictions and analytical results. Since throughout the paper we provide numerical evidence supporting the predictions of the theory, here is a brief summary: the models and parameters, the initial states, and the chosen measurements. The complete details of the numerics are given in the methods section. We focus on 1D spin-1/2 models with nearest-neighbor interactions and magnetic fields. Our Hamiltonians are all translationally invariant so that the parameters of the model are the nearest neighbor interaction constants $\vec{J}$ and the magnetic field $\vec{B}$. We investigated 7 different models, spanning through quantum integrable and quantum chaotic and with system sizes up to $N=20$. The explicit values are given in table \ref{table:models}. We looked at 5 different initial states $\ket{\psi_0(\theta_m)}$, with $m\in\{1,5,10,15,20\}$. These are all tensor product states with N\'eel order, but aligned along different axes. For $m=1$ the initial state is aligned along $z$ while for $m=20$ it is aligned along $x$. The rest are aligned along an axis forming an angle $\theta = \frac{m-1}{19} \frac{\pi}{2}$ with the $z$ axis, in the $z-x$ plane. We highlight that in the parameter regime considered these systems do not have weak coupling for either one-site or two-site subsystems according to any definition we are aware of. This is discussed in detail in the Supplementary Information. For all Hamiltonian models and all initial states, we studied 6 classes of observables, three one-body and three two-body: $\sigma_i^{x},\sigma_i^{y},\sigma_i^z,\sigma_i^x\sigma^x_{i+1},\sigma_i^y\sigma^y_{i+1},\sigma_i^z\sigma^z_{i+1}$, with $i=1,\ldots, N$.

\subsubsection{Dynamical Relaxation to Max Observable Entropy}
Since Observable Statistical Mechanics is based on the Maximum Observable Entropy Principle, we checked that the observable entropy exhibits relaxation towards a high-entropy stationary state. We indeed observe this behavior for almost models, initial states and observables considered. An example is given in Fig.~\ref{fig:entropy_X_forall_m_model1_no_inset}, while additional plots are included in the Supplementary Information.
\begin{figure}
    \centering
    \includegraphics[width=\linewidth]{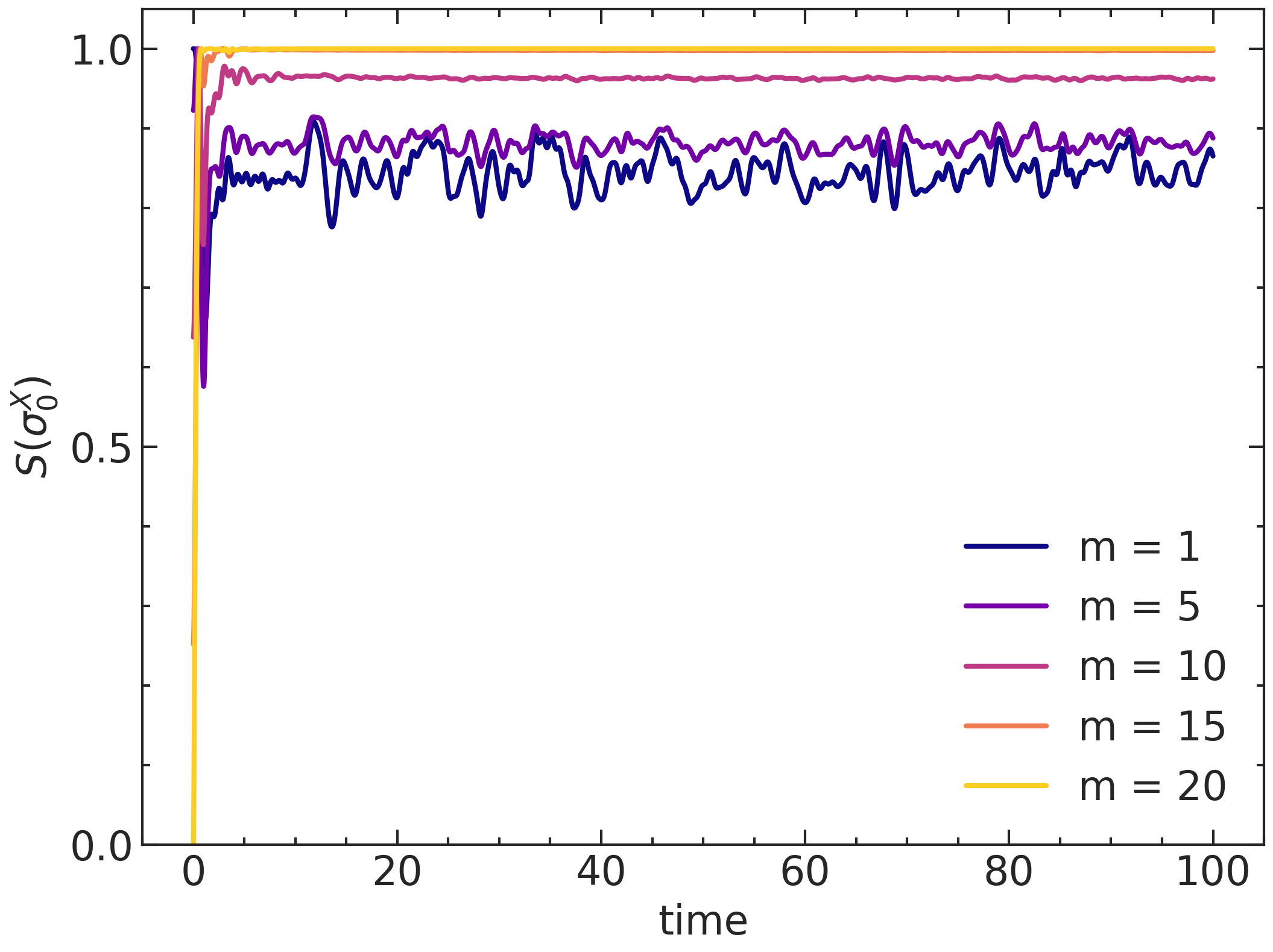}
    \caption{Time evolution of the Shannon entropy $S(A)$ for model $J=(0,0,1) \, \& \, B=(0.9045, 0, 0.8090)$, observable $\sigma^x_0$ and all the initial states considered (labeled by $m$), for $N=20$. Note the constrained maximum can be lower than the absolute maximum. In fact, this plot has been chosen to highlight the above point, and in other models, for the same observable, entropy equilibrates around the absolute maximum for every initial state (see the Supplementary Information).}
    \label{fig:entropy_X_forall_m_model1_no_inset}
\end{figure}
This shows numerical evidence of a dynamical tendency to evolve and settle on high observable entropy configurations---a necessary prerequisite for the validity of the theory. It also supports the heuristic arguments presented before that the stationary value of the entropy, computed from the diagonal ensemble distribution $p_j^{DE}$, can be characterized via the Maximum Observable Entropy Principle. These findings are supported by the independent study that emerged after this work \cite{meier_emergence_2024}, typicality arguments \cite{popescu_entanglement_2006, malabarba_quantum_2014} and the effective dynamics study in \cite{strasberg_classicality_2023}.

\subsubsection{Equilibrium Equations} \label{subsubsec:EEs}
We proceed by introducing the Equilibrium Equations of Observable Statistical Mechanics, together with their general solution and an ensuing prediction. 

The constrained maximization of $\tilde{S}_A$ can be solved using the Lagrange multipliers techniques \cite{reichl_modern_2016, anza_information-theoretic_2017}. As discussed above, we keep only the normalization of probabilities and average energy constraints. After various algebraic manipulations, which are detailed in the Methods section \ref{subsec:EEs_derivation}, we find that when a highly-degenerate observable $A$ satisfies the MOEP its distribution $p_j^{DE}$ will obey the following Equilibrium Equation:
\begin{gather}
    - p_j^{DE} \log p_j^{DE} + p_j^{DE} \log d_j = \lambda_A p_j^{DE} + \beta_A R_j^{DE} \label{eqn:EE2}~,
\end{gather}
where
\begin{align}
    &R_j^{DE} \coloneqq \Tr(A_j H \rho_{DE})= \sum_n |c_n|^2 E_n \left[ A_j\right]_{nn}~,
\end{align}
and $\lambda_A,\beta_A$ are the Lagrange multipliers associated with the normalization and average energy constraint, respectively. Despite Eq.~\eqref{eqn:EE2} being highly non-linear and exhibiting an implicit dependence on $p_j^{DE}$, there is one and only one form of the equilibrium probability distribution that is compatible with a maximum entropy characterization: the exponential distribution. This follows from Gibbs' inequality \cite{jaynes_brandeis_1989}. Properly parameterized for future convenience, this leads to 
\begin{align}
&p_j^{DE}=\frac{d_je^{-\beta_A \varepsilon_j}}{\mathcal{Z}_A} &&\mathcal{Z}_A = \sum_j d_j e^{-\beta_A \varepsilon_j}\label{eqn:exact_sol}
\end{align}
where $\varepsilon_j$ is an observable-specific quantity with the dimension of energy such that $\sum_j p_j^{DE} \varepsilon_j = E$, and $\mathcal{Z}_A$ is a measurement-specific partition function. Its meaning is understood and it will be discussed in subsection \ref{subsubsec:understanding_epsj}. 

\subsubsection{A special solution: Hamiltonian Unbiased Observables} \label{subsubsec:HUOs}

While the most general solution of the Equilibrium Equations, Eq.~\eqref{eqn:exact_sol}, is given here for the first time, a special case had been previously studied in \cite{anza_information-theoretic_2017, anza_eigenstate_2018}: $p_j = \frac{d_j}{D}$. This is the equilibrium distribution of a large class of observables, the Hamiltonian Unbiased Observables (HUOs), for which $\varepsilon_j = E \; \forall j$. These are observables which are diagonal in a basis that is unbiased \cite{durt_mutually_2010, bandyopadhyay_new_2002, wehner_entropic_2010} with the energy basis: $\langle E_n | j,s \rangle = \frac{e^{i\theta_{js}^n}}{\sqrt{D}}$. Such a basis is, accordingly, called a Hamiltonian Unbiased Basis (HUB). Physically, these observables do not possess any information about the energy; in fact, in \ref{subsubsec:MI_HUOs} we show their mutual information with the Hamiltonian is zero. HUOs are a very useful model to understand thermalization, for three reasons. First, it has been proven analytically that the ETH holds for all of them \cite{anza_information-theoretic_2017}, so they will generically thermalize under standard assumptions. Second, in a statistically precise sense (Haar measure), most observables are expected to be quite close to being HUOs. This was proven in Ref.~\cite{anza_eigenstate_2018}, along with other statements clarifying their physical relevance. Third, using them we can determine extensive sets of thermal observables in MBL systems. Indeed, while it is true that MBL systems escape quantum statistical mechanics in the standard sense, there are several local observables that still exhibit MOEP, even in the localized phase \cite{anza_pure_2018, anza_logarithmic_2020}. \\

While HUOs capture core aspects of observable thermalization, they have one major drawback: they are insensitive to the overall energy scale of the system. This is because their equilibrium mutual information with the energy vanishes identically. We do not expect this to hold exactly for physical observables which, at equilibrium, do exhibit a smooth dependence on the energy scale of the system. Nevertheless, we believe the core mechanism to be approximately correct. In other words, we expect a non-vanishing but very small equilibrium mutual information between the observable and energy distributions. This is important because it guarantees it is sufficient to impose the average energy constraint when maximizing the entropy.

\subsubsection{A novel prediction} \label{subsubsec:Rjpj}
\begin{figure}
    \centering
    \includegraphics[width=\linewidth]{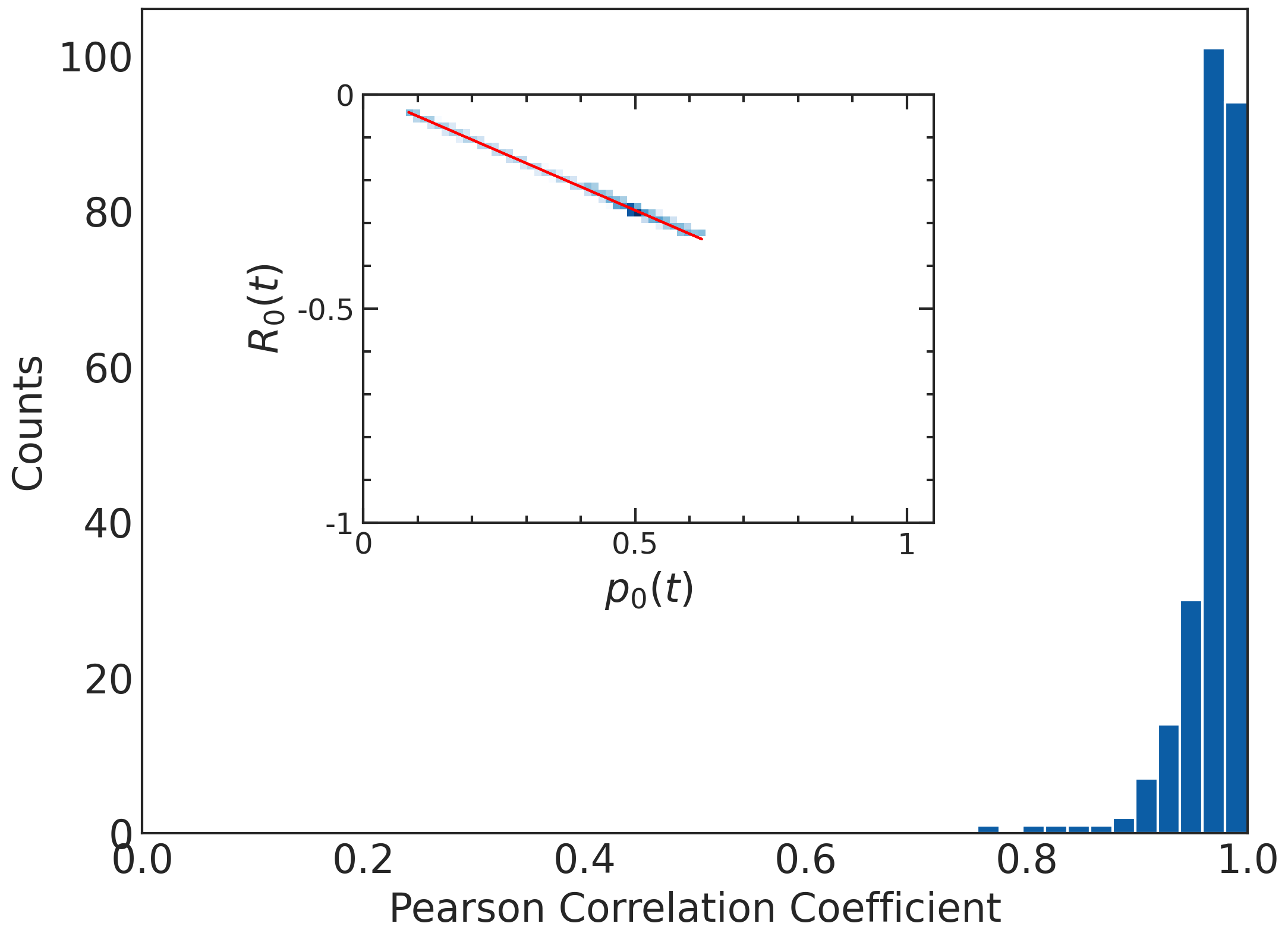}
    \caption{Here we provide support for the existence of a linear relationship between $R_j$ and $p_j$. The main plot shows the histogram of the (absolute value of the) Pearson correlation coefficient of the sets $\{p_j(t), R_j(t)\}$ for every (independent) eigenvalue of all observables considered, for all Hamiltonian models and initial states. Note that sets essentially corresponding to points or constant lines (i.e., cases where either standard deviation $\Delta p_j$ or $\Delta R_j$ is very small) have been excluded, as their Pearson coefficient would incorrectly indicate non-linearity. In the inset, we give as an example the 2D histogram of the time-implicit plot of $(p_0(t), R_0(t))$ of the observable $\sigma^y_0$ for model $J=(0,0,1) \, \& \, B=(0.9045, 0, 0.8090)$ and initial state corresponding to $m=10$. Note we are using a logarithmic color scale, and the red line is a linear fit. We rescaled the extensive quantity $R_j(t) \to R_j(t)/N$ to be able to compare it among different system sizes. In both plots, the system size is $N=20$.}\label{fig:linearity_pjRj}
\end{figure}

By plugging the general solution Eq.~\eqref{eqn:exact_sol} into the Equilibrium Equation Eq.~\eqref{eqn:EE2}, we find a relation that defines $\varepsilon_j$ at equilibrium, namely,
\begin{equation} \label{eqn:desired_form_Rj}
    R_j^{DE} = \varepsilon_j p_j^{DE}~.
\end{equation}
Together, eqs.\eqref{eqn:EE2},\eqref{eqn:exact_sol} and \eqref{eqn:desired_form_Rj} constitute our first result, summarized as follows: The MOEP provides solvable equilibrium equations whose solution predicts that the stationary $p_j^{DE}$ is of the Gibbs (exponential family) form where the notion of energy of the $j-$th outcome is played by $\varepsilon_j = R_j^{DE}/p_j^{DE}$. In Section \ref{subsec:bound_Rj} we show that $\varepsilon_j \in \mathcal{I}_{mc}(E, \Delta E)$. Moreover, to confirm that Eq.~\eqref{eqn:desired_form_Rj} is approximately true even if we are not exactly on the diagonal ensemble, we numerically test its validity. To see the dynamically-emergent linearity, we use a time-implicit plot $(p_j(t),R_j(t))$, where $R_j(t)=\bra{\psi_t} \frac{HA_j+A_jH}{2}\ket{\psi_t}$ is the time-dependent version of $R_j^{DE}$ such that $\overline{R_j(t)}= R_j^{DE}$. For each such plot, we compute the (absolute value of the) Pearson correlation coefficient, which quantifies the linear correlation between the two sets $\{p_j(t)\}$ and $\{R_j(t)\}$, with a value of $1$ corresponding to perfect linearity. In figure \ref{fig:linearity_pjRj}, we give the histogram of all Pearson coefficients, computed numerically: the vast majority of values is very close to 1. Moreover, the inset shows an example of the 2D histogram of a plot of $R_j(t)$ against $p_j(t)$, where the linear behavior is also evident. Thus, we have strong evidence that supports the validity of Eq.~\eqref{eqn:desired_form_Rj} even if we only have apparent equilibration. This in turn supports the use of the exponential solution Eq.~\eqref{eqn:exact_sol} for $p_j^{DE}$.

\subsubsection{An observable-specific notion of energy.} \label{subsubsec:understanding_epsj}

To obtain a complete solution, we need to fix the value of the Lagrange multiplier $\beta_A$ by using the energy constraint:  
\begin{equation} \label{eqn:ave_lam}
    \beta_A \,\, : \,\, \sum_j p_j^{DE} \varepsilon_j =- \frac{\partial \log \mathcal{Z}_A}{\partial \beta_A}= E .
\end{equation}
Since the solution to the constrained optimization problem must belong to the exponential family, much like Boltzmann's distribution, the $\left\{\varepsilon_j\right\}_{j=1}^{n_A}$ play the role of the energy spectrum for the chosen observable $A$. What is, then, the physical interpretation of $\varepsilon_j$? And, can we compute it without knowledge of the energy eigenstates? To answer these important questions we first write explicitly $\varepsilon_j$:
\begin{equation} \label{eqn:epsjeq_ratio_Rjpj_DE}
    \varepsilon_j = \frac{R_j^{DE}}{p_j^{DE}} = \sum_n \frac{ |c_n|^2 [A_j]_{nn} E_n}{\sum_m |c_m|^2 [A_j]_{mm}} \\
\end{equation}
We now consider that the stationary state $\rho_{DE}$ is the same as the post-measurement state obtained when we measure energy: $\rho_{DE}$ is the diagonal part of $\ketbra{\psi_t}{\psi_t}$ in the energy basis. We then build the equilibrium joint probability distribution of the energy eigenvalue and the $j$-th measurement outcome by specifying the order of measurements as ``energy first''. This is motivated by the fact that we are interested in stationary values: $p_{eq}(E_n,j):= |c_n|^2 [A_j]_{nn}$. Then, using Bayes' theorem, we have 
\begin{equation}
p_{eq}(E_n \vert j)  \coloneqq \frac{p_{eq}(E_n,j)}{\sum_n p_{eq}(E_n,j)} =  \frac{ |c_n|^2 [A_j]_{nn}}{\sum_m |c_m|^2 [A_j]_{mm}}.
\end{equation}
The expectation value of the energy with the conditional distribution $p_{eq}(E_n | j)$ is then $\varepsilon_j$: 
\begin{equation}
\sum_n p_{eq}(E_n \vert j)E_n = \varepsilon_j  
\end{equation}
Hence, $\varepsilon_j$ is the conditional expectation of the energy at fixed measurement outcome $j$. Or, equivalently, $\varepsilon_j$ is the fraction of average energy stored, at equilibrium, in the eigenspace $\mathcal{H}_j$. This is our second result: \emph{$p_j^{DE}$ is Boltzmann's distribution of the conditional energies $\varepsilon_j$ of the $j-$th measurement outcome}. Uncovering $\varepsilon_j$ as a novel notion of measurement-specific energy paints a compelling picture for the emergence of statistical mechanics (and thermodynamics) of observables, rather than states.

\subsubsection {Computing \texorpdfstring{$\varepsilon_j$}{εⱼ}} \label{subsubsec:computing_epsj}
\begin{figure}[ht]
    \centering
    \includegraphics[width=\linewidth]{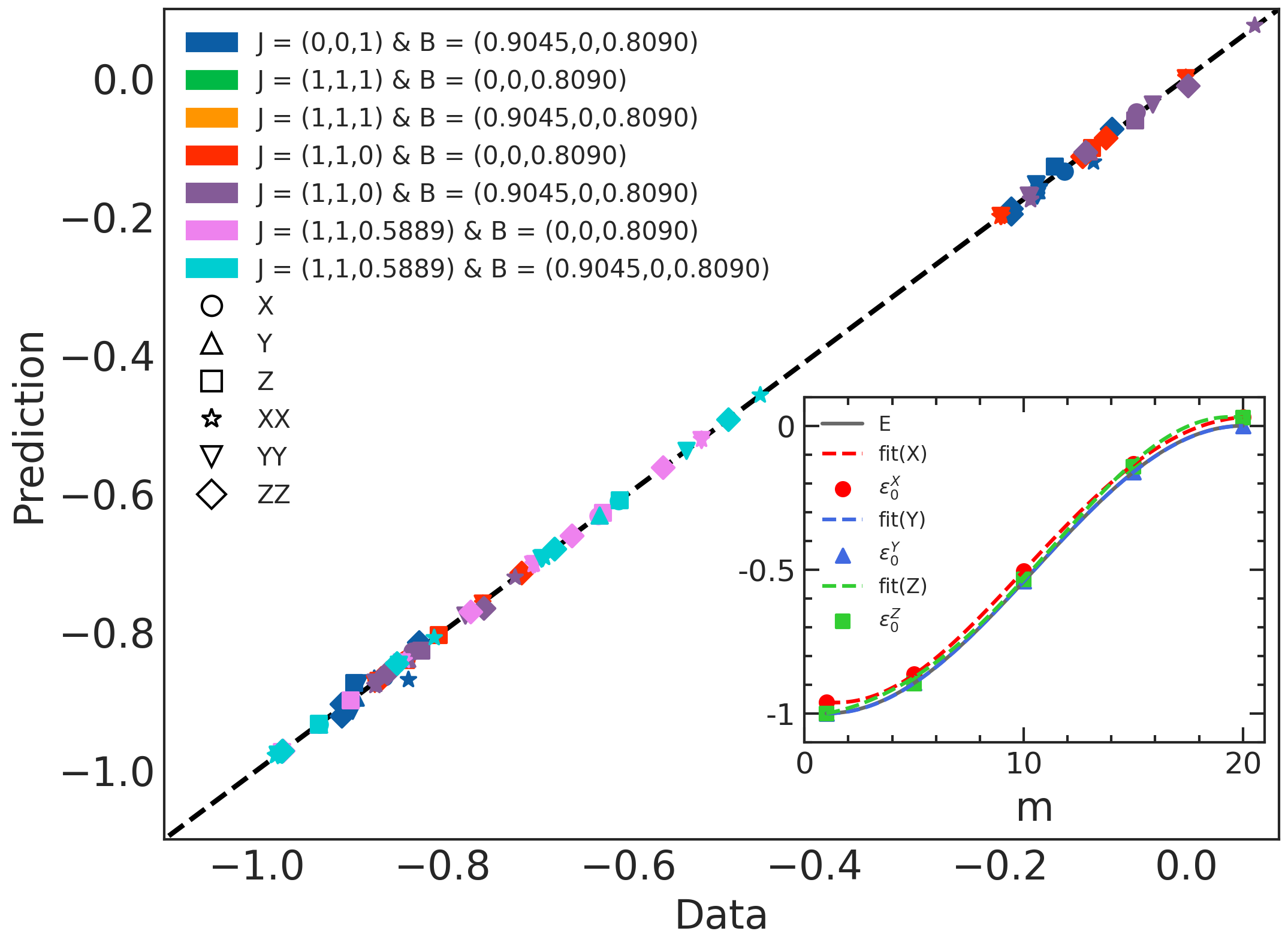}
    \caption{Implicit plot of our prediction for $\varepsilon_j^A(\theta_m)$ given the fit obtained using Eq.~\eqref{eqn:epsj_theta} against the numerical data, for $N=20$. The plot includes $\varepsilon_j$'s for all (independent) eigenvalues of all observables, for all models and initial states. The black dashed line has been added for visual aid and represents a perfect prediction. We only include comparisons with the $\varepsilon_j^{data}$ that were \textit{not} used to determine the coefficients $\gamma_j, \eta_j, \chi_j$. In the inset, we show as an example a plot of $\varepsilon_j^A(\theta_m)$ and $E(\theta_m)$ against the initial state, parametrized by $m$ (as described in subsection \ref{subsec:numerics}), for the eigenvalue $j=0$ of one-body observables, model $J=(0,0,1) \, \& \, B=(0.9045, 0, 0.8090)$ and $N=20$. To make the plot more readable, we use the simplified notation $X \equiv \sigma^x_0$, $Y \equiv \sigma^y_0$, and $Z \equiv \sigma^z_0$. The colourful dashed lines are the fits to the data points given the function Eq.~\eqref{eqn:epsj_theta}. Note that we are using analytical expressions for the energy average $E(\theta_m)$ and standard deviation $\Delta E(\theta_m)$, shown in the Supplementary Information. Note also that we divided by $N$ the extensive quantities $\varepsilon_j, E, \Delta E$ to be able to compare them among different system sizes.}
    \label{fig:prediction_vs_data}
\end{figure}

While understanding $\varepsilon_j$ completes the theory at the conceptual level, having a way to compute it is paramount to leverage its predictive power. Here we provide a practical recipe based on the following analytical argument. First, $\varepsilon_j$ must be approximately linear in the average energy: $\varepsilon_j = \gamma_j E + \chi_j$, at fixed $p_j^{DE}$. This can be seen from the constraint equation \eqref{eqn:ave_lam}, or from Eq.~\eqref{eqn:desired_form_Rj}. However, at the most fundamental level, these equations are actually implicit and non-linear in $\varepsilon_j$, so $\varepsilon_j$ can actually vary due to fluctuations. Nevertheless, for thermodynamically stable initial states (see Reimann's work \cite{reimann_foundation_2008} and the discussion in section \ref{subsec:heuristics}), we can see that $\varepsilon_j$ can take values within the microcanonical window, $\varepsilon_j \in [E - \Delta E/2, E + \Delta E/2]$, with small fluctuations allowed. We then land on the empirically linear form
\begin{equation} \label{eqn:epsj_theta}
    \varepsilon_j = \gamma_j E + \eta_j \Delta E+ \chi_j.
\end{equation}
It is important to highlight that the contribution from $\Delta E$ is expected to be negligible in most cases. This is especially true in the macroscopic limit $N \to \infty$, where $\Delta E / E \to 0$. However, in some cases, it must be taken into account, including when comparing situations in which we have the same average energy, but different fluctuations. In those cases, $\eta_j$ will play a more important role in correctly estimating $\varepsilon_j$. Thus, $\varepsilon_j$'s value is entirely determined by \emph{at most} three coefficients $\gamma_j,\eta_j,\chi_j$ and, most often, $\gamma_j$ and $\chi_j$ are sufficient. For example, since we are exploring isolated quantum systems with initial states $\ket{\psi_0(\theta)}$  numerical knowledge of two or three values of $\varepsilon_j$ is sufficient to determine $\varepsilon_j$ for all initial states. This can be seen in Fig.~\ref{fig:prediction_vs_data}, where we do precisely this: we use 2-3 data points to fix the coefficients $\gamma_j,\eta_j,\chi_j$ and then use Eq.~\eqref{eqn:epsj_theta} to predict the remaining $\varepsilon_j$. We see that the predictions match the data extremely well.
The inset shows $\varepsilon_j^A(\theta_m)$ and $E(\theta_m)$ plotted against the initial state in one case. The fits, found using Eq.~\eqref{eqn:epsj_theta} and exact analytically expressions for $E(\theta_m)$ and $\Delta E(\theta_m)$ (given in the Supplementary Information), are clearly excellent. Additional data in support of these statements are shown in the Supplementary Information.

\subsubsection{Predictive power} \label{subsubsec:testing_theory}

\begin{figure}
    \centering
    \includegraphics[width=\linewidth]{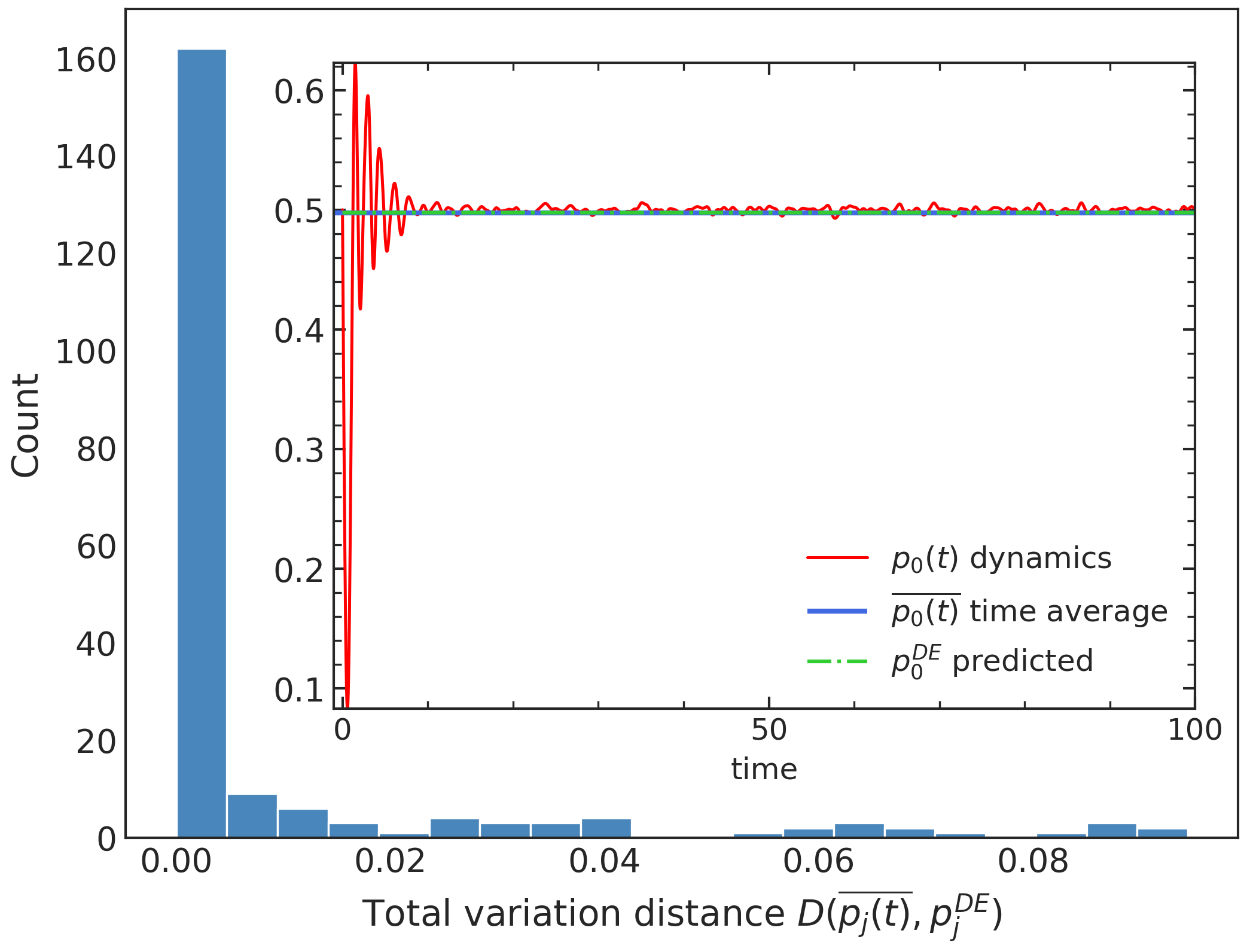}
    \caption{Histogram of the total variation distance $D(\overline{p_j(t)}, p_j^{DE})$ for all observables, models and initial states considered, for $N=20$. The inset shows as an example the time evolution of the exact probability distribution $p_0(t)$, together with its time average and our equilibrium prediction, for the observable $\sigma^y_0$, model $J=(0,0,1) \, \& \, B=(0.9045, 0, 0.8090)$, initial state $m=10$ and $N=20$. Note that the x-axis has been limited to small values of the total variation distance.}
    \label{fig:goodness_prediction}
\end{figure}

We now test the predictive power of Observable Statistical Mechanics by putting the theoretical framework that has been laid out so far into work. For all observables, initial states, and Hamiltonian models, we find the Lagrange multiplier $\beta_A$ that solves the energy constraint $\sum_j p_j^{DE}\varepsilon_j = E$ and use it to compute the analytical prediction Eq.~\eqref{eqn:exact_sol}. 
Then, we compare it with the time-average $\overline{p_j(t)}^T\coloneqq \frac{1}{T}\int_0^T p_j(t) dt$ obtained for sufficiently large values of $T$ from the numerics. The comparison is made using the total variation distance between two probability distributions, i.e., $D(p, q) := \frac{1}{2} \sum_j |p_j - q_j|$, and the results are summarized in figure \ref{fig:goodness_prediction}. We observe a remarkable agreement with the theory, with $D(\overline{p_j(t)}, p_j^{DE}) \leq 10^{-9}$ for all one-body observables, models and initial states considered. For two-body observables, the distance $D(\overline{p_j(t)}, p_j^{DE})$ is generally higher than for the one-body, but overall there is an extremely good agreement with the theory's predictions, with $\sim 80\%$ of values under $10^{-2}$. The data is shown as a histogram in the main plot of Fig.~\ref{fig:goodness_prediction}, which includes both one and two-body observables. The inset instead gives an example of the time evolution of the exact probability distribution $p_j(t)$. We see that after a short transient the probability exhibits small fluctuations around its equilibrium value, which agrees extremely well with the predictions of Observable Statistical Mechanics.

\section{Discussion}\label{sec:discussion}

We established Observable Statistical Mechanics: a new way to predict the stationary behavior of interacting, many-body, isolated quantum systems. Rather than attempting to justify the emergence of thermal states from dynamical arguments, Observable Statistical Mechanics replaces the notion of thermal equilibrium with the Maximum Observable Entropy Principle: an information-theoretic principle stating that stationary measurement statistics are well-described by the distribution maximizing the entropy $\tilde{S}(A)$. The predictions of Observable Statistical Mechanics have been compared with numerical experiments, showing a remarkable capability to predict the stationary behavior of one-body and two-body observables in 7 different spin-1/2 Hamiltonians, spanning both chaotic and integrable models, and with 5 different initial states. A few comments are in order. \\

\begin{figure}
    \centering
    \includegraphics[width=\linewidth]{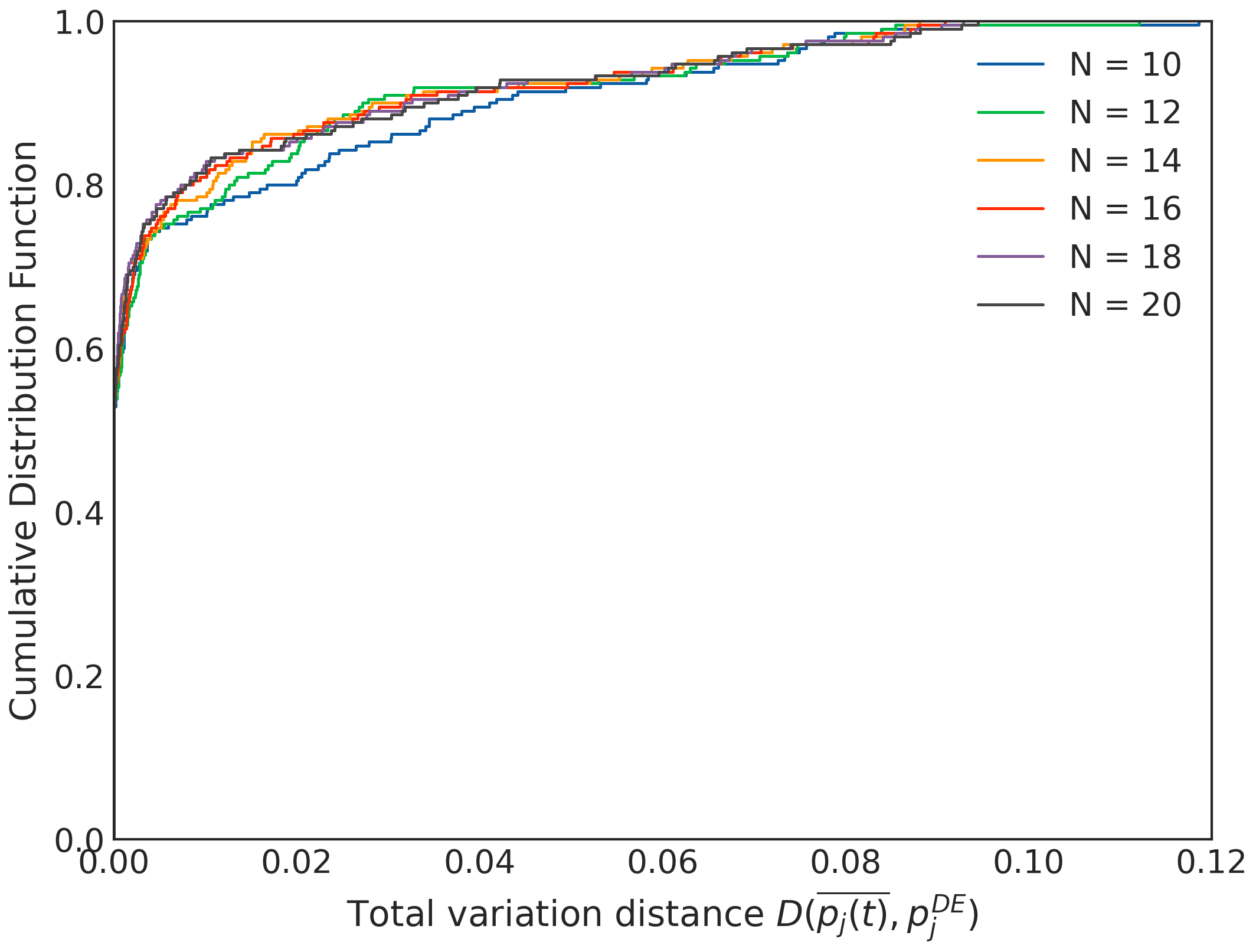}
    \caption{Comparison of the goodness of our prediction among system sizes ranging from $N=10$ to $N=20$. The figure shows the cumulative distribution function of the total variation distance. We see we have similarly good agreement for all system sizes, slightly improving with increasing $N$. Note that the x-axis has been limited to small values of the total variation distance.}
    \label{fig:goodness_prediction_with_L}
\end{figure}

First, Observable Statistical Mechanics goes beyond standard Equilibrium Statistical Mechanics. By not relying on the weak coupling assumption, we are able to make prediction about coarse observables in the diagonal ensemble. In the Supplementary Information we show that, for all non-integrable models and initial states we have considered, and for both one-site and two-site subsystems, the coupling does not appear to be sufficiently weak to justify the use of the Gibbs ensemble in these parameter regimes. This conclusion holds for all three definitions of weak coupling we have found in the literature. Moreover, our predictions in principle apply also to highly degenerate subspaces of non-local observables, so it would be interesting to study them in future work. Further, the accuracy of our predictions does not depend strongly on the system size. Indeed, this can be seen in Fig.~\ref{fig:goodness_prediction_with_L}, where we show our predictions hold for system sizes ranging from $N=10$ to $N=20$. This is a great practical advantage since the exponential growth of the Hilbert space dimension is often the true bottleneck.

Second, the mechanism for the emergence of a maximum observable entropy principle in isolated many-body quantum dynamics is analytically understood in information-theoretic terms, using Holevo's bound: measuring realistic, coarse, observables is akin to sending information through a high-noise quantum communication channel. This is absent in ordinary quantum statistical mechanics, and it is an element of novelty with respect to the standard characterization of quantum integrable vs. quantum chaotic systems, which is often used to discuss the nature of equilibrium. Observable Statistical Mechanics considers the dominant role played by measurements and aims at providing an accurate estimate of the stationary equilibrium distribution of a set of measurements, irrespective of the integrable or chaotic nature of the system. \\

Third, it is reasonable to wonder when the framework can fail. Firstly, we note that the observable must equilibrate, as otherwise it would not even make sense to talk about stationarity. A quantity that is well-known to play a crucial role in equilibration studies is the effective dimension $d_\textrm{eff} := \frac{1}{\sum_n |c_n|^4}$, which quantifies the number of energy eigenstates that are significantly populated by the initial state. Its importance relative to the equilibration of observable entropies has also been highlighted recently in Ref.~\cite{meier_emergence_2024}. We note that the R\'enyi-2 entropy of the energy distribution $\{|c_n|^2\}$ is precisely $S_2(H) = \log d_\textrm{eff}$, and that $S(H) \geq S_2(H)$. Hence, having an effective dimension that grows exponentially with the system size $d_\textrm{eff} \sim D^{\frac{k_E}{\log d}}$ guarantees both equilibration and an extensive entropy $S(H) \gtrsim k_E N$, which in turn implies that $\tilde{S}(A) \geq S(H)$ is also extensive at equilibrium. \\
Further, the predictions might also fail when $\frac{I(H,A)}{S(H)} \in \mathcal{O}(1)$. This could be due an energy distribution with low entropy $S(H) \ll N$, which happens, for example, when the initial state is the superposition of only a few energy eigenstates, thus leading to lack of equilibration as discussed above. Alternatively, we could have a measurement channel with high communication capacity: $I(H,A) \sim N$. This happens, for example, for small systems. In those cases, we can build experiments that can probe an entire basis, thus accessing an exponentially large number of microscopic details. While in this case it might still be possible, by chance, to observe a stationary measurement statistics that satisfies MOEP, we are not guaranteed that it will happen. \\

Eventually, by moving the focus from states to measurement statistics, this work opens up new research avenues into the foundations of equilibration and thermalization. We mention a few. Firstly, the observable-specific notion of energy that we introduced, the $\varepsilon_j$, deserves further study. Given the crucial role it plays in predicting the stationary distribution, the empirical method provided to calculate it should be better understood, and it would be beneficial to find a fully analytical way of computing this quantity. Tentatively, our result supports the hypothesis advanced by Strasberg and collaborators in \cite{strasberg_classicality_2023} that an effectively Markov dynamics satisfying detailed balance can emerge from a coherent dynamics. It also advances it by providing a simple prediction for what the notion of energy involved in the transition rates and detailed balance should be: the $\varepsilon_j$. Specifically, assuming the validity of an effective master equation, as advocated in \cite{strasberg_classicality_2023}, our analytical results and numerical evidence support the idea that the transition rates $\gamma_{jk}$ involved in the master equation should satisfy the following form of detailed balance:
\begin{align}
    \frac{\gamma_{jk}}{\gamma_{kj}} = e^{-\beta_A(\varepsilon_k - \varepsilon_j)}
\end{align}
Unfortunately, the mathematical complexity of the problem did not allow the authors in \cite{strasberg_classicality_2023} to cast their result in the form of a rigorous theorem. Therefore, we draw the connection at the physical level and only conjecture the connection at the mathematical, rigorous, level.

Secondly, this new notion of energy suggests the possibility of quantum thermodynamics based on measurement statistics rather than density matrices. Indeed, thanks to our understanding of $\varepsilon_j$ we can define an \emph{observable free energy} $F_A$ via $\mathcal{Z}_A \eqqcolon e^{-\beta_A F_A}$. Then, summing both sides of the Equilibrium Equation \eqref{eqn:EE2} over $j$, we find that at equilibrium we have
\begin{align}
    F_A = E - T_A \tilde{S}_A~,
\end{align}
with the observable temperature $T_A \coloneqq \beta_A^{-1}$. The emergence of this equilibrium relation, which is essentially a form of energy conservation, is a prediction of Observable Statistical Mechanics. However, to have a complete picture we need a kinetic, or operational, interpretation of the observable-specific thermodynamic quantities $T_A$ and $F_A$. This is currently being developed and will be reported in future work. Finally, there has been significant interest lately on the effect of non-commuting charges (non-Abelian symmetries) on equilibration and thermalization \cite{majidy_noncommuting_2023}. It would be interesting to explore the predictive power of Observable Statistical Mechanics in this case, and whether the presence of these more complex symmetries significantly affects the premises of the theory.

\section{Methods}\label{sec:methods}

\subsection{Validity of the Maximum Entropy Principle} \label{subsubsec:validity_maxent}

In subsection \ref{subsec:heuristics} we gave two heuristic arguments to support the use of the Maximum Observable Entropy Principle: one concerning the equilibrium mutual information between the observable and energy distributions, the other involving the extensivity of the entropy of the observable's complete eigenbasis. We now give the full version of these arguments. 

\subsubsection{Defining joint, conditional and marginal probabilities} \label{subsubsec:def_prob_entropy_mutual_info}
The study of equilibrium concerns the property of the diagonal ensemble $\rho_{DE}:= \sum_n |c_n|^2 \vert E_n \rangle \langle E_n \vert$. This is the diagonal part of the time-dependent pure state $\ket{\psi_t}$ in the energy basis and also the post-measurement state if we were to measure the energy of the system. Using the measurement interpretation we can compute the joint distribution of the energy and our observable by assuming the order ``energy first''. This is because we are interested in equilibrium properties. Hence, we have that the energy marginal distribution is $p_{DE}(E_n) = |c_n|^2$, and the conditional is $p_{DE}(j,s\vert E_n) = \vert \langle E_n\vert j,s \rangle\vert^2$. Eventually, this leads to the following joint probability distribution at equilibrium: $p_{DE}(E_n; j,s) = |c_n|^2 \vert \langle E_n\vert j,s \rangle\vert^2$. From this we obtain the marginal $p_{DE}(j,s)$ by summing over the energy variable: $p_{DE}(j,s) = \sum_n |c_n|^2 \vert \langle E_n\vert j,s \rangle\vert^2$.

The conditional probability of observing the $n$-th eigenvalues at equilibrium, given $(j,s)$ can be computed by means of Bayes' theorem:
\begin{equation}
    p_{DE}(E_n \vert j,s) = \frac{p_{DE}(E_n; j,s)}{p_{DE}(j,s)} = \frac{|c_n|^2 \vert \langle E_n\vert j,s \rangle\vert^2}{\sum_m |c_m|^2 \vert \langle E_m\vert j,s \rangle\vert^2}.
\end{equation}
Since the index $s$ accounts for the degeneracy of the observables under study, it is also useful to compute the equilibrium probability distributions that concern only the measurement outcomes. Hence, we have the additional probabilities $p_{DE}(E_n; j) = |c_n|^2 [A_j]_{nn}$, $p_{DE}(j) = \sum_{n}|c_n|^2 [A_j]_{nn}$, $p_{DE}(E_n \vert j) = \frac{p_{DE}(E_n ; j)}{p_{DE}(j)} = \frac{|c_n|^2[A_j]_{nn}}{\sum_m |c_m|^2[A_j]_{mm}}$.

\subsubsection{Zero Equilibrium Mutual Information for HUBs and HUOs} \label{subsubsec:MI_HUOs}
What's needed to make accurate predictions about equilibrium is in the energy eigenstates. However, our argument is that highly degenerate observables will be able to retain very little information about these. Certainly, not as many details as the energy eigenstates themselves. Therefore, instead of using all conserved quantities, knowledge of the average energy should be sufficient to make accurate predictions. \\
To support this line of thinking we compute two variants of the classical equilibrium mutual information between energy and observable. The first one $\tilde{I}_{DE}(H,A)$ is about the mutual information between the energy and a full basis that diagonalizes our observables. The second one $I_{DE}(H,A)$ is about the mutual information between the energy and the observable, disregarding the detailed information about the degeneracy. \\
We now compute these quantities for HUBs and HUOs, to check if the core intuition is correct. Recall a basis $\{ \vert j,s \rangle \}_{j=1,s=1}^{j=n_A,s=d_j}$ is a HUB if $\langle E_n \vert j,s \rangle = \frac{e^{i\theta_{js}^n}}{\sqrt{D}}$, and an observable is HUO if it is diagonalized by a HUB \cite{anza_information-theoretic_2017}. This leads to $p_{DE}(j,s) = p_{DE}(j,s\vert E_n) = \frac{1}{D}$, $p_{DE}(E_n) = p_{DE}(E_n \vert j,s) = |c_n|^2$ and, similarly, for the coarse-grained distribution of outcomes of a HUO we find $p_{DE}(j) = p_{DE}(j\vert E_n) = \frac{d_j}{D}$, $p(E_n) = p_{DE}(E_n \vert j) = |c_n|^2$.

Clearly, the measurement basis carries no information about the Hamiltonian basis, and viceversa: the conditionals are equal to the marginals. In other words, the two variables are independent. It's easy to see that the mutual information $I(X, Y) := \sum p(X, Y) \log \left( \frac{p(X, Y)}{p(X) p(Y)} \right)$ is zero if and only if $X$ and $Y$ are independent random variables, so this implies
\begin{align}
& \tilde{I}_{DE}^{HUB}(H,A) = I_{DE}^{HUO}(H,A)=0.
\end{align}

\subsubsection{Equilibrium Mutual Information for Highly Degenerate Observables} \label{subsubsec:MI_highly_deg_obs}
We now generalize these arguments about the mutual information to general highly degenerate observables. To do so, we make use of the aforementioned Theorem 1 in \cite{anza_eigenstate_2018}, which says that for all subspaces such that $d_j(d_j-1)\geq D+1$ there is a measurement basis such that $\left\vert \langle E_n \vert j,s\rangle \right\vert^2 = \frac{[A_j]_{nn}}{d_j}$. This leads to probability distributions that weigh equally all the different observable degenerate states:
\begin{align}
& p_{DE}(E_n;j,s) = \frac{|c_n|^2 [A_j]_{nn}}{d_j} = \frac{p_{DE}(E_n,j)}{d_j}\\
&p_{DE}(j,s) = \frac{\sum_n |c_n|^2 [A_j]_{nn}}{d_j}=\frac{p_{DE}(j)}{d_j} \\
&p_{DE}(j,s\vert E_n) = \frac{[A_j]_{nn}}{d_j} = \frac{p_{DE}(j\vert E_n)}{d_j}\\
&p_{DE}(E_n \vert j,s) = \frac{[A_j]_{nn}|c_n|^2}{\sum_{m}[A_j]_{mm}|c_m|^2} = p_{DE}(E_n \vert j)
\end{align}
This implies that $\tilde{I}_{DE}(H,A) = I_{DE}(H,A), $ as can be easily verified by direct calculation. \\

\paragraph*{Negligible Mutual Information in the Thermodynamic Limit.}
Note that the quantities $\tilde{I}_{DE}(H,A)$ and $I_{DE}(H,A)$ have very different scales. In fact, the mutual information is upper-bounded by the entropy marginals, namely $\tilde{I}\leq \min\{\tilde{S}(A), S(H)\}$ and $I\leq \min\{ S(A), S(H) \}$. However, while $\tilde{S}(A)\leq \log D$ we have that $S(A)\leq \log n_A \ll \log D$. Nevertheless, since in this case these two mutual informations are equal, we just need to consider the scaling behaviour of $S(A)$ and $S(H)$. The latter is expected to be extensive $S(H) \sim N \sim \log D$. For the former, we note that in most experimental apparata our measurements will have at best $n_A = N^k$ different outcomes, for some integer $k \sim \mathcal{O}(1)$, so $S(A) \lesssim \log N$. In fact, we have already discussed how few-body observables have $n_A = d^{N_S}$ and extensive sums of local observables such as the global magnetization have $n_A \sim N$. Hence, as argued previously, in many cases observables have only few outcomes and thus act as a noisy channels, i.e., 
\begin{equation}
    \frac{\tilde{I}(H, A)}{S(H)} = \frac{I(H, A)}{S(H)} \leq \frac{S(A)}{S(H)} \leq \frac{\log n_A}{S(H)} \ll 1.
\end{equation}
Similarly, following Balz and Reimann \cite{balz_equilibration_2016}, no realistic measurement can give us more than 20 relevant digits. Hence $n_A < 10^{20}$, which is independent of $N$. The conclusion is  unchanged: $\frac{\tilde{I}_{DE}(H,A)}{S(H)} \ll 1$ for large $N$.

As we approach the thermodynamic limit, in a macroscopically stable systems highly degenerate observables will not carry significant information about the distribution of energies. Therefore, when setting up our maximum entropy principle, the simple use of the average energy constraint should be sufficient. \\

\paragraph*{Finite-size argument.}
The previous argument shows that, when considering larger and larger systems, we will eventually be justified to consider that our observables will be diagonalized by a basis that carries a negligible amount of information about the distribution of energies at equilibrium. However, this leaves open the question of what is the appropriate scale in which this becomes relevant. We now argue that the core argument can actually work rather well independent on the system size. \\
To see this, note that, 
\begin{align}
    & \tilde{I}_{DE} = I_{DE} = \sum_n |c_n|^2\left[S(A) - S(A \vert E_n) \right],
\end{align}
with $S(A\vert E_n) = - \sum_j p_{DE}(j\vert E_n) \log p_{DE}(j\vert E_n)$ the entropy of the distribution of the measurement outcomes conditioned on the $n$-th energy eigenspace. Suppose $|c_n|^2$ is a narrow energy distribution, concentrated around the mean value, and $p_{DE}(A \vert E_n)$ does not fluctuate too much, which is similar to assuming the diagonal part of the ETH. Then, for all energy eigenvalues within a microcanonical window $E_n \in [E - \Delta E/2 E+ \Delta E/2]$ we will have $p_{DE}(j \vert E_n) \approx \sum_m |c_m|^2 p_{DE}(j \vert E_m) = p_{DE}(j)$, which leads to $S(A) \approx \sum_n |c_n|^2 S(A\vert E_n)$ and eventually to $I_{DE} \ll S(H)$.

\subsubsection{Extensive Entropy Argument} \label{subsubsec:extensive_entropy}
The previous argument regarding the equilibrium mutual information justifies our use of only two constraints in the entropy maximization problem for HUOs and, more generally, for highly degenerate observables. We shall now motivate the use of a maximum entropy principle for such observables in the first place, by showing $\tilde{S}(A)$ is extensive in these cases. We already argued previously that this is true in general for any observable at equilibrium if $S(H)$ itself is extensive. This is because the Schur-Horn theorem and the Schur concavity of the Shannon entropy imply that $\tilde{S}(A) \geq S(H)$ at equilibrium. Here we will show that for highly degenerate observables we can find another lower bound that is stronger in some cases, for instance for few-body observables. The following chain of inequalities holds at equilibrium for observables for which $\exists j: \max_s p_{js} = \max_{j's} p_{j's} \land d_j(d_j-1) \geq D+1$:
\begin{equation}
    \begin{aligned}
    \tilde{S}(A) &:= -\sum_{js} p_{js} \log p_{js} \geq -\log\left( \max_{j,s} p_{js}\right) = \\
                 &= -\log\left( \max_j \frac{p_j}{d_j} \right) \geq \min_j \log d_j + S_\infty\left(\{p_j\}\right)
    \end{aligned}
\end{equation}
where $S_\infty\left(\{p_j\}\right) := \min_j \left( -\log p_j\right)$ is the min-entropy. In the second line, we have used Theorem 1 in \cite{anza_eigenstate_2018}, which states that when $d_j(d_j-1) \geq D+1$ there is always a basis $\{ \vert j,s \rangle \}$ such that $\vert \langle j,s\vert E_n\rangle\vert^2 = \frac{[A_j]_{nn}}{d_j}$. This implies $p_{js} = \frac{p_j}{d_j}$ at equilibrium. \\
Let us look at the two terms on the right hand side of the final inequality individually. Since $S_{\infty}(\{p_j\}) \in [0, \log n_{A}]$, the second term would give us at most a contribution of $\mathcal{O}(\log N)$ for realistic measurements, as previously discussed for the Shannon entropy $S(A)$ in \ref{subsubsec:MI_highly_deg_obs}. The first term depends on the type of highly degenerate observable we are considering. For instance, suppose we have an observable whose eigensubspaces have constant degeneracy and for which $d_j(d_j-1)\geq D - 1$ holds. In the regime $d_j \gg 1$, we have $d_j \sim D^{\kappa_j} \Rightarrow \log d_j \sim (\kappa_j \log d) N$ with $\frac{1}{2} \leq \kappa_j \leq 1$. Observables that have support on $N_S$ sites with $n_A = d^{N_S}$ outcomes and that obey these conditions will have $\log d_j = (N-N_S) \log d$. However, we note that if instead an observable has at least one eigenspace with very low degeneracy, then this is not a good bound. For instance, for the total magnetization we have $\min_j \log d_j = 0$. \\
This argument agrees well with the scaling we obtain from the relation
\begin{equation} \label{eqn:observational_entropy}
    \tilde{S}(A) = S(A) + \sum_j p_{DE}(j) \log d_j,
\end{equation}
which can be found by applying the aforementioned Theorem 1 in \cite{anza_eigenstate_2018} to the definition of $\tilde{S}(A)$ at equilibrium, assuming the conditions of the theorem apply to all eigenspaces. As stated previously, $S(A) \in [0, \log n_A]$ will scale at most logarithmically in the system size $\sim \log N$, while the second term will scale like $\sim \left(\sum_j p_j \kappa_j \right) N$, with the sum being of $\mathcal{O}(1)$ for an equilibrium probability distribution. \\
An extensive entropy is the hallmark of statistical mechanics. This argument provides a rigorous basis to the intuition that, due to high degeneracy, equilibrium measurement statistics will have a highly entropic diagonalizing basis. In turn, this justifies the idea that accurate predictions at equilibrium can be made using the Maximum Observable Entropy Principle.

\subsection{Derivation of the Equilibrium Equations} \label{subsec:EEs_derivation}
The original Equilibrium Equations were derived in \cite{anza_information-theoretic_2017} by maximizing $S(A)$ under the constraints of state normalization and fixed average energy. Here we sketch the derivation of the Equilibrium Equations obtained by maximizing $\tilde{S}(A)$ under the same constraints. The derivation is analogous to that in \cite{anza_information-theoretic_2017}, so we refer the reader to that paper for the details. \\
We consider a generic mixed state $\rho(t) = \sum_k q_k \vert \psi_k(t)\rangle\langle \psi_k(t)\vert$ and the entropy $\tilde{S}(A)$. Moreover, we define the quantity $D_{js}^k := \langle j,s\vert \psi_k\rangle$ and its complex conjugate $\overline{D_{js}^k}$. The two constraints are
\begin{align}
    \mathcal{C}_N &:= \Tr(\rho) - 1 \mbeq 0 \\
    \mathcal{C}_E &:= \Tr(\rho H) - E \mbeq 0.
\end{align}
We can then define the auxiliary function $\Lambda_A [\rho; \lambda_N, \beta_A] := \tilde{S}(A) + \lambda_N \mathcal{C}_N + \beta_A \mathcal{C}_E$, where $\lambda_N$ and $\beta_A$ are respectively, the Lagrange multipliers for the normalization and average energy constraint. Taking derivatives of this function with respect to $D_{js}^k$ and $\overline{D_{js}^k}$, and performing some algebraic manipulations, gives us
\begin{gather}
    \Tr\big(\rho \left[ A_{js}, H \right]\big) \eqeq 0 \label{eqn:EE1-methods}\\
    - p_{js} \log p_{js} \eqeq \lambda_A p_{js} + \beta_A R_{js} \label{eqn:EE2-methods} 
\end{gather}
where $\lambda_A := (1 + \lambda_N)$, $A_{js} := \vert j,s \rangle\langle j,s \vert$ and 
\begin{equation}
    R_{js} := \Tr\big(\rho \{ A_{js}, H \}\big) = \operatorname{Cov}(A_{js}, H) + p_{js} E
\end{equation}
with the anticommutator $\left\{ X,Y\right\} \coloneqq \frac{XY+YX}{2}$ and the symmetrized covariance $\operatorname{Cov}(X,Y)\coloneqq \langle\{X, Y\}\rangle - \expval{X}\expval{Y}$. We note that the object $R_{js}$ is an inner product between the operators $A_{js}$ and $H$. \\

The first Equilibrium Equation, Eq.~\eqref{eqn:EE1-methods}, is about dynamical equilibration, i.e., it states that the distribution must be invariant under the unitary dynamics generated by $H$. Indeed, using von Neumann's equation one has $i\hbar\frac{\partial}{\partial t} p_{js} \eqeq 0$. The second Equilibrium Equation characterizes the shape of $p_{js}^{DE}$. For example, if we sum over $j,s$ we can see that the entropy of the equilibrium distribution has a thermodynamic flavor in the sense that it has a linear relation with the average energy:
\begin{equation} \label{eqn:equilibrium_obs_entropy}
    \tilde{S}(A) \eqeq \log \mathcal{Z}_A + \beta_A E,
\end{equation}
where $\beta_A$ plays the role of a measurement-specific inverse temperature and, calling $\log \mathcal{Z}_A \coloneqq \lambda_A$, $\mathcal{Z}_A$ is a measurement-specific partition function. \\

From Eq.~\eqref{eqn:EE2-methods}, we can obtain Eq.~\eqref{eqn:EE2} by using Theorem 1 in \cite{anza_eigenstate_2018}, which implies $p^{DE}_{js} = \frac{p_j^{DE}}{d_j}$ and $R^{DE}_{js} = \frac{R_j^{DE}}{d_j}$ at equilibrium. This is equivalent to assigning to each degeneracy within the $j$-th eigenspace the same probability $\frac{1}{d_j}$, at equilibrium. This is related to the definition of the Observational Entropy \cite{safranek_quantum_2019, strasberg_first_2021, safranek_brief_2021}; see also Eq.~\eqref{eqn:observational_entropy}. And, indeed, this provides a physically meaningful mechanism for the emergence of this thermodynamic entropy --- it corresponds to the Shannon entropy of the complete eigenbasis of an equilibrating and highly degenerate observable (all of whose eigenspaces obey $d_j(d_j-1)\geq D+1$), computed on the diagonal ensemble.

\subsection{Bound on \texorpdfstring{$R_j$}{Rⱼ}} \label{subsec:bound_Rj}
Within a microcanonical energy window $\mathcal{I}_{mc} := \left[ E - \frac{\Delta E}{2}, E + \frac{\Delta E}{2} \right]$, $R^{DE}_j \approx \sum_{E_n \in \mathcal{I}_{mc}} |c_n|^2 E_n [A_j]_{nn}$. Using $E_n \in \mathcal{I}_{mc},$ we find the following straightforward bound 
 \begin{equation}\label{eqn:bound_Rj}
    \bigg(1 - \frac{\Delta E}{2E}\bigg)p^{DE}_j \lesssim \frac{R^{DE}_j}{E} \lesssim \bigg(1 + \frac{\Delta E}{2E}\bigg)p^{DE}_j    
 \end{equation}
This implies that $\varepsilon_j := \frac{R^{DE}_j}{p^{DE}_j} \in \mathcal{I}_{mc}$. Moreover, in the thermodynamic limit and for thermodynamically stable states, we also expect $\Delta E/E \ll 1$, so as we increase the system size $\varepsilon_j$ will tend to the average energy.

\subsection{Numerics} \label{subsec:numerics}
We now provide more details regarding the numerical simulations we have conducted. We have considered $7$ one-dimensional spin-$1/2$ models described by Hamiltonians of the form
\begin{equation} \label{eqn:general_hamiltonian}
    H = \sum_{i=0}^{N-1} \sum_{\alpha = x, y, z} \big( J^\alpha \sigma^\alpha_i \sigma^\alpha_{i+1} + B^\alpha \sigma^\alpha_i \big),
\end{equation}
where $i$ is an index that runs over the $N$ lattice sites, and $\sigma^\alpha_i$ ($\alpha=x,y,z$) represents the Pauli operator with Pauli matrix $\sigma^\alpha$ acting on the lattice site $i$, i.e.
\begin{equation} \label{eqn:local_pauli}
 \sigma^\alpha_i \coloneqq \mathbb{I}^{\otimes i} \otimes \sigma^\alpha \otimes \mathbb{I}^{\otimes(N-1-i)}.
\end{equation}
We use periodic boundary conditions, so that $\sigma^\alpha_N = \sigma^\alpha_0$. We denote the $7$ models considered by their interaction coefficients $J = (J^x, J^y, J^z)$ and their magnetic field coefficients $B = (B^x, B^y, B^z)$, which are shown in table \ref{table:models}.

\begin{table}[h]
    \centering
    \begin{tabular}{c|c|c|c}
         Model&  $(J^x, J^y, J^z)$& $(B^x, B^y, B^z)$ & Integrable? \\
         \hline
         Ising + LT &  (0, 0, 1)& (0.9045, 0, 0.8090)& No \cite{kim_testing_2014}\\
         XXX + L &  (1, 1, 1)& (0, 0, 0.8090)& Yes \cite{franchini_introduction_2017} \\
         XXX + LT &  (1, 1, 1)& (0.9045, 0, 0.8090) & Yes \cite{franchini_introduction_2017} \\
         XX + L &  (1, 1, 0)& (0, 0, 0.8090) & Yes \cite{franchini_introduction_2017}\\
         XX + LT &  (1, 1, 0)& (0.9045, 0, 0.8090) & No \cite{dmitriev_gap_2002} \\
         XXZ + L &  (1, 1, 0.5889)& (0, 0, 0.8090) & Yes \cite{franchini_introduction_2017} \\
         XXZ + LT &  (1, 1, 0.5889)& (0.9045, 0, 0.8090) & No \cite{dmitriev_gap_2002} \\
         \hline
    \end{tabular}
\caption{The $7$ Hamiltonian models considered in the numerical simulations, classified in terms of their interaction and magnetic field coefficients. Of the models chosen, $3$ are non-integrable and $4$ are integrable. ``L" and ``LT" respectively refer to the presence of only a longitudinal field, or both longitudinal and transverse fields.}
\label{table:models}
\end{table}

These models range from strongly non-integrable, such as in the case $J = (0,0,1)$ \& $B = (0.9045, 0, 0.8090)$, which is guaranteed to have such behaviour by \cite{kim_testing_2014}, to fully integrable models such as in the case $J = (1,1,0)$ \& $ B = (0, 0, 0.8090)$. \\
To obtain the state dynamics, we use Trotterization, but we are guaranteed the error scales only linearly with the system size due to the form of the Hamiltonian (see the discussion in the Supplementary Information). We evolve the system up to time $t_f=100$ with a timestep $dt = 10^{-3}$, resulting in $100,000$ time points. \\
We have considered (even) system sizes from $N=10$ up to $N=20$ spins, and a class of initial states given by 
\begin{equation} \label{eqn:psi_theta_m}
    \ket{\psi_0(\theta_m)}=R_y(\theta_m)^{\otimes N}\ket{01\ldots01},
\end{equation}
where $R_y(\theta) = \exp(-iY\theta/2)$ is the rotation operator along the $y$ axis and $\theta_m = \frac{m-1}{19}\frac{\pi}{2}$. These states interpolate between the antiferromagnetic state along the $z$ direction ($\theta_1 = 0$) and the one along the $x$ direction ($\theta_{20} = \frac{\pi}{2}$).
Note that the extremes are states of zero Shannon entropy for observables such as $\sigma^z_i$ and $\sigma^x_i$ respectively, thus maximally out of equilibrium for these observables. \\
In the Supplementary Information we argue that, according to all three definitions of weak coupling we have identified in the literature, the interaction between one-site and two-site subsystems and their effective environments \textit{cannot} be neglected for all models, initial states and system sizes considered in our numerics. In particular, we show that the operator norm of the interaction Hamiltonian $\| H_\textrm{int} \|$ is not negligible compared to either the subsystem's Hamiltonian's norm $\| H_S \|$, the subsystem's minimal energy gap $\Delta E_S$ or the global energy standard deviation $\Delta E$. \\
Throughout this study, we consider one-body observables $\sigma^\alpha_i$ of the form given in Eq.~\eqref{eqn:local_pauli}, and two-body observables of the form
\begin{equation} \label{eqn:two-body_obs}
 \sigma^\alpha_i \sigma^\alpha_{i+1} = \mathbb{I}^{\otimes i} \otimes \sigma^\alpha \otimes \sigma^{\alpha} \otimes \mathbb{I}^{\otimes(N-2-i)}.
\end{equation}
Their coarse-grained eigen-projectors are respectively
\begin{gather} \label{eqn:coarse-grained_projectors}
    A_j^{\alpha,i} = \frac{\mathbb{I} + (-)^{j} \sigma^\alpha_i}{2} \\
    A_{jk}^{\alpha,i} = A_j^{\alpha,i} A_k^{\alpha,i+1}
\end{gather}
with $j,k \in \{0,1\}$. Note this means that for two-body observables we have four distinct eigenvalues, i.e., $\{00,01,10,11\}$. Throughout the paper, we present only plots for the lattice site $i=0$, as there is no difference between sites due to the translation invariance of the models. \\
To predict the equilibrium distribution $p_j \propto e^{-\beta_A \varepsilon_j}$, we first need to compute $\varepsilon_j$ and then $\beta_A$. For the former, we first extract 3 data points from the numerics, either via a linear fit on the data $\{p_j(t), R_j(t)\}$ or by taking the ratio of the means $\overline{R_j(t)}$ and $\overline{p_j(t)}$. Then, we use Eq.~\eqref{eqn:epsj_theta} to compute the remaining $\varepsilon_j$ for all initial states. For the latter, $\beta_A$, for one-body (i.e., binary ($j=0,1$)) observables we have an exact analytical solution
\begin{equation} \label{eqn:lambdaE_binary_obs}
    \beta_A = \frac{1}{\delta\varepsilon} \operatorname{arctanh}\bigg( \frac{\Bar{\varepsilon} - E}{\delta\varepsilon} \bigg) = \frac{1}{\varepsilon_1 - \varepsilon_0} \ln\bigg( \frac{\varepsilon_1 - E}{E - \varepsilon_0} \bigg),
\end{equation}
where $\Bar{\varepsilon} \coloneqq \frac{\varepsilon_1 + \varepsilon_0}{2}$ and $\delta\varepsilon \coloneqq \frac{\varepsilon_1 - \varepsilon_0}{2}$. For two-body observables, instead, we find the value of $\beta_A$ by numerically optimizing Eq.~\eqref{eqn:ave_lam} with respect to it.
\\
Finally, all extensive quantities at equilibrium have been rescaled by $1/N$ to ensure that we can compare them among different system sizes.

\begin{acknowledgments}
L.S. thanks the ``Angelo Della Riccia" Foundation for their continued support, and is grateful to Luis Pedro Garc\'ia Pintos, Samuel Slezak and Zo\"e Holmes for helpful discussions. F.A. acknowledges support from the Templeton World Charity Foundation under grant TWCF0336. F.A. would like to thank J.P. Crutchfield and C. Jarzynski for discussions about the dynamical emergence of thermal equilibrium. V.V. thanks the Gordon and Betty Moore Foundation and the Templeton Foundation for supporting his research. The authors would like to acknowledge the use of the University of Oxford Advanced Research Computing (ARC) facility in carrying out this work \cite{richards_university_2015}, and they are grateful to Tristan Farrow for providing them with access to this computing cluster. Finally, the authors acknowledge the use of Qibo \cite{efthymiou_qibo_2022, efthymiou_quantum_2022} in the simulations.
\end{acknowledgments}

\section*{Author Contributions}
F.A. conceived the idea and L.S. contributed to its development. L.S. also contributed to the analytical calculations, the numerical analysis and the preparation of the final manuscript. F.A. also directed the project, and contributed to the analytical calculations and the preparation of the final manuscript. A.A. worked on the numerical simulations, and V.V. contributed to the development of the project.

\section*{Competing Interests}
The authors declare no competing interests.

\section*{Data Availability}
All data are available upon reasonable request to the authors.


%

\clearpage
\appendix
\onecolumngrid

\section{Additional numerical data} \label{sec:additional_numerics}
In this appendix, we provide additional numerical evidence supporting Observable Statistical Mechanics. We investigated $7$ different Hamiltonian models, with system sizes from $10$ to $20$ (even only), initialized in $5$ different initial states, and then studied the dynamics of the full probability distribution of $6$ different observables. This amounts to $1260$ long-time dynamics of  probability distributions. Due to the large-scale nature of the numerical experiments, we cannot provide here figures for all data studied. Therefore, we only show selected additional figures to support and complement the plots shown in the main text. All data are available upon reasonable request to the authors.

\subsection{Numerics} \label{subsec:numerics_app}
Before we show the additional data, here is a brief summary of the details of the numerics, for the reader's convenience. We have considered $7$ one-dimensional spin-$1/2$ models described by Hamiltonians of the form
\begin{equation} \label{eqn_app:general_hamiltonian}
    H = \sum_{i=0}^{N-1} \sum_{\alpha = x, y, z} \big( J^\alpha \sigma^\alpha_i \sigma^\alpha_{i+1} + B^\alpha \sigma^\alpha_i \big),
\end{equation}
where $i$ runs over the $N$ lattice sites, and $\sigma^\alpha_i$ ($\alpha=x,y,z$) represents the Pauli operator with Pauli matrix $\sigma^\alpha$ acting on the lattice site $i$, i.e.
\begin{equation} \label{eqn_app:local_pauli}
 \sigma^\alpha_i \coloneqq \mathbb{I}^{\otimes i} \otimes \sigma^\alpha \otimes \mathbb{I}^{\otimes(N-1-i)}.
\end{equation}
We use periodic boundary conditions, so that $\sigma^\alpha_N = \sigma^\alpha_0$. The $7$ models considered have different  interaction coefficients $J = (J^x, J^y, J^z)$ and magnetic field coefficients $B = (B^x, B^y, B^z)$, which are shown in the following table.

\begin{table}[h]
    \centering
    \begin{tabular}{c|c|c|c}
         Model&  $(J^x, J^y, J^z)$& $(B^x, B^y, B^z)$ & Integrable? \\
         \hline
         Ising + LT &  (0, 0, 1)& (0.9045, 0, 0.8090)& No \cite{kim_testing_2014_app}\\
         XXX + L &  (1, 1, 1)& (0, 0, 0.8090)& Yes \cite{franchini_introduction_2017_app} \\
         XXX + LT &  (1, 1, 1)& (0.9045, 0, 0.8090) & Yes \cite{franchini_introduction_2017_app} \\
         XX + L &  (1, 1, 0)& (0, 0, 0.8090) & Yes \cite{franchini_introduction_2017_app}\\
         XX + LT &  (1, 1, 0)& (0.9045, 0, 0.8090) & No \cite{dmitriev_gap_2002_app} \\
         XXZ + L &  (1, 1, 0.5889)& (0, 0, 0.8090) & Yes \cite{franchini_introduction_2017_app} \\
         XXZ + LT &  (1, 1, 0.5889)& (0.9045, 0, 0.8090) & No \cite{dmitriev_gap_2002_app} \\
         \hline
    \end{tabular}
\caption{The $7$ Hamiltonian models considered in the numerical simulations, classified in terms of their interaction and magnetic field coefficients. Of the models chosen, $3$ are non-integrable and $4$ are integrable. ``L" and ``LT" respectively refer to the presence of only a longitudinal field or both longitudinal and transverse fields.}
\label{tab_app:models_params}
\end{table}

We evolve the system up to time $t_f=100$ with a timestep $dt = 10^{-3}$, resulting in $100,000$ time points. \\
We have considered (even) system sizes from $N=10$ up to $N=20$ spins, and a class of initial states given by 
\begin{equation} \label{eqn_app:psi_theta_m}
    \ket{\psi_0(\theta_m)}=R_y(\theta_m)^{\otimes N}\ket{01\ldots01},
\end{equation}
where $R_y(\theta) = \exp(-iY\theta/2)$ is the rotation operator along the $y$ axis and $\theta_m = \frac{m-1}{19}\frac{\pi}{2}$. These states interpolate between the antiferromagnetic state along the $z$ direction ($\theta_1 = 0$) and the one along the $x$ direction ($\theta_{20} = \frac{\pi}{2}$).
Note that the extremes are states of zero Shannon entropy for observables such as $\sigma^z_i$ and $\sigma^x_i$ respectively, thus maximally out of equilibrium for these observables.
We considered one-body observables $\sigma^\alpha_i$ of the form given in Eq.~\eqref{eqn_app:local_pauli}, and two-body observables of the form
\begin{equation} 
 \sigma^\alpha_i \sigma^\alpha_{i+1} = \mathbb{I}^{\otimes i} \otimes \sigma^\alpha \otimes \sigma^{\alpha} \otimes \mathbb{I}^{\otimes(N-2-i)}.
\end{equation}

\subsection{Additional data} \label{subsec:additional_data}
\begin{figure}[ht]
    \centering
    \includegraphics[width=\linewidth]{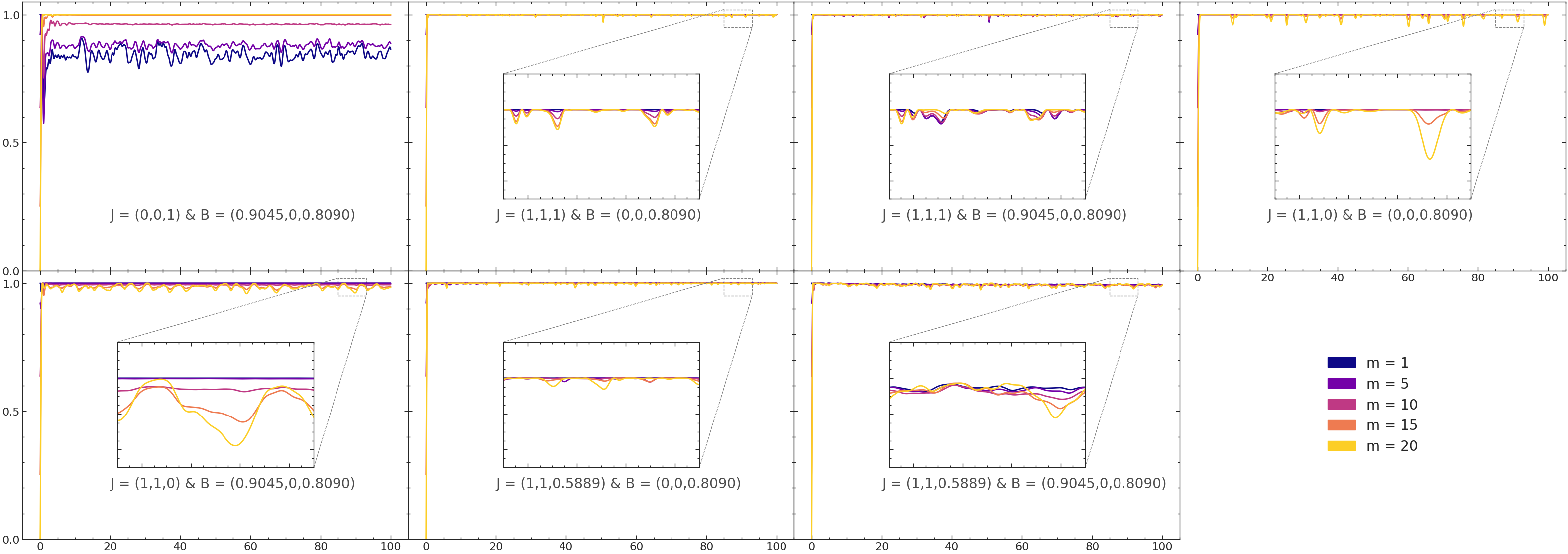}
    \caption{Here we show the time evolution of the Shannon entropy $S(A)$ for the observable $\sigma^x_0$ for all the models and initial states $m$ considered, with $N=20$. Each subplot corresponds to one of the Hamiltonians models we used. For the vast majority of other cases, we observe the same qualitative behavior, i.e., the entropy equilibrates around the maximum.}
    \label{fig:entropy_X_forall_mH}
\end{figure}

\begin{figure}[ht]
    \centering
    \includegraphics[width=0.9\linewidth]{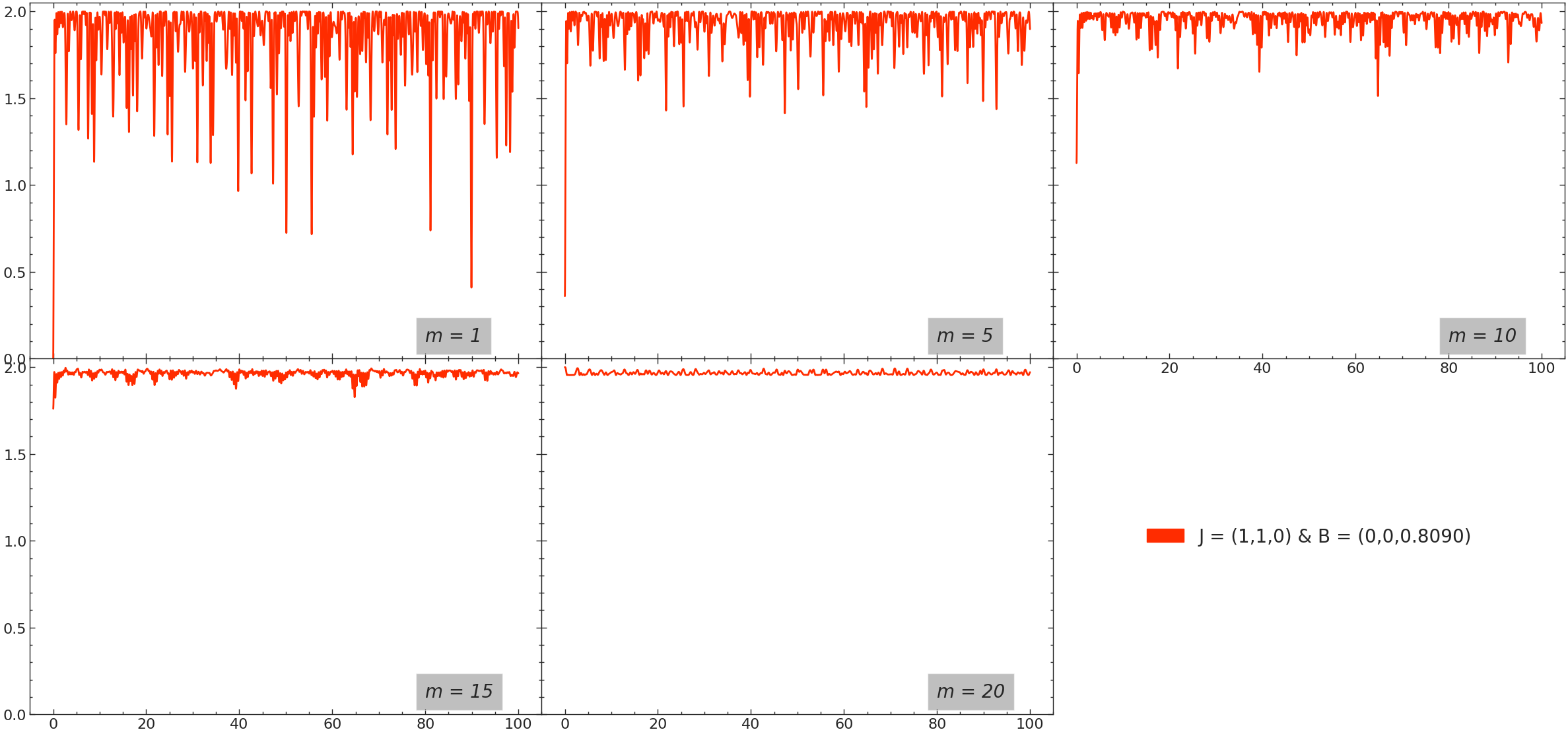}
    \caption{Here we give a rare example of the entropy $S(A)$ not equilibrating. This plot corresponds to the observable $\sigma^{z}_0 \sigma^{z}_1$ for the model $J=(1,1,0) \, \& \, B=(0, 0, 0.8090)$ and for all initial states $m$ considered, with $N=20$. Each subplot corresponds to one of the initial states we used. We note that whether the entropy equilibrates or not for this model depends on the initial state. The particular behavior we observe in this case might be due to the fact that this Hamiltonian is not only integrable, but can also be mapped to a non-interacting fermion model via a Jordan-Wigner transformation \protect\cite{jordan_uber_1928_app, franchini_introduction_2017_app}.}
    \label{fig:entropy_ZZ_forall_m_model4}
\end{figure}

In Fig.~\ref{fig:entropy_X_forall_mH} we give further evidence beyond what shown in the main text that indeed entropy is maximized, irrespective of the model considered. The same qualitative behavior can be seen for almost all other observables, models and initial states we investigated. Nevertheless, in Fig.~\ref{fig:entropy_ZZ_forall_m_model4} we show a rare case of the entropy not reaching equilibrium. It is interesting to observe that this behavior emerges only for a specific combination of observables, models and initial states, and indeed the dependence on the initial state can clearly be seen in the figure. That is to say, having only knowledge of the model is insufficient to predict the equilibrium features of the entropy; one needs to specify the observable and the initial state too.

\begin{figure}[ht]
    \centering
    \subfloat[$\varepsilon_j^A(\theta_m)$ and $E(\theta_m)$ against $m$, for the eigenvalue $j=00$ of two-body observables for the model $J=(0,0,1) \, \& \, B=(0.9045, 0, 0.8090)$, with $N=20$. To make the plot more readable, we use the simplified notation $XX \equiv \sigma^x_0 \sigma^x_1$, $YY \equiv \sigma^y_0 \sigma^y_1$, and $ZZ \equiv \sigma^z_0 \sigma^z_1$. \label{fig:epsj_vs_m_extra_2-body}]{\includegraphics[width=0.45\textwidth]{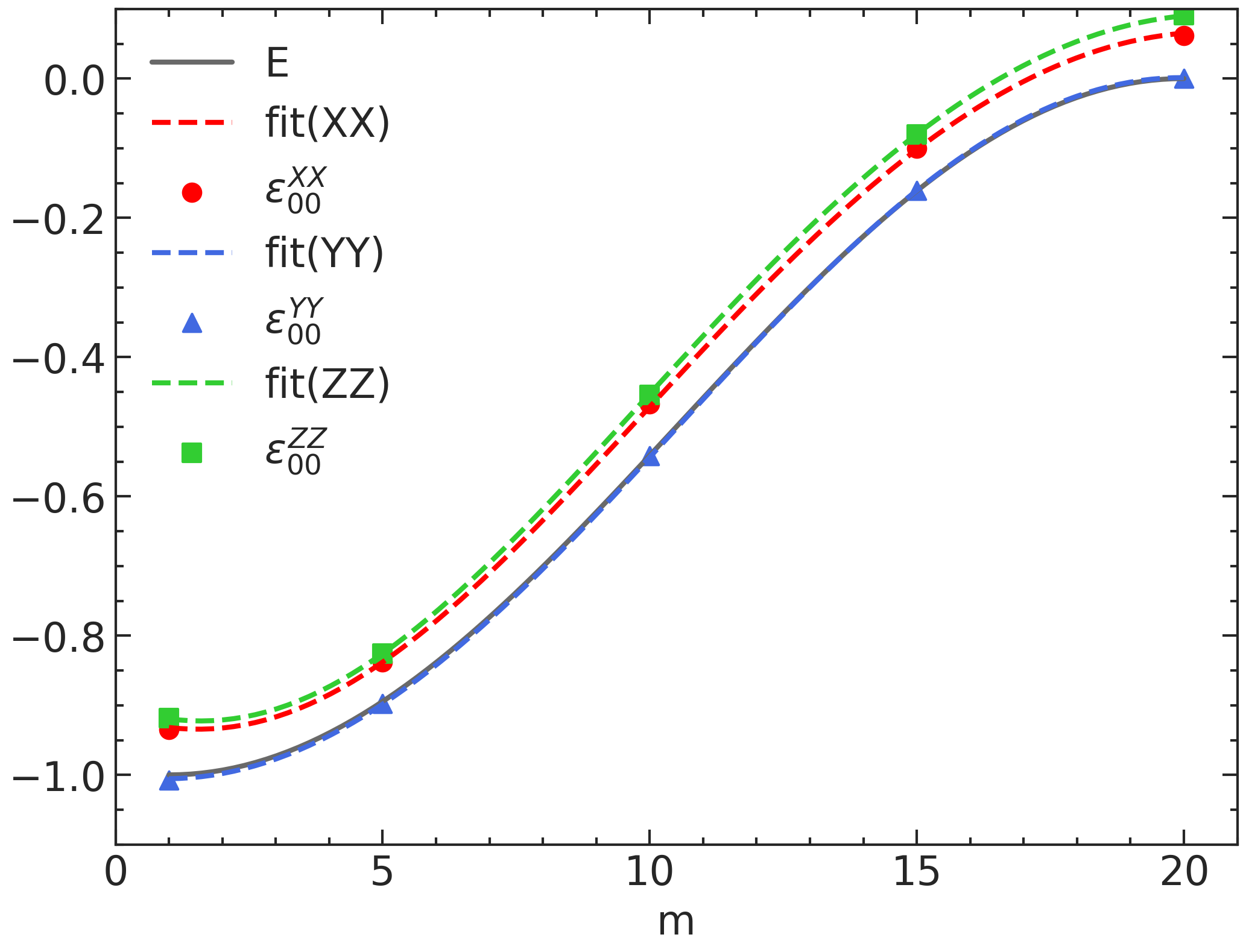}} \hspace{1cm}
    \subfloat[$\varepsilon_j^A(\theta_m)$ and $E(\theta_m)$ against $m$, for the eigenvalue $j=0$ of one-body observables for the model $J=(1,1,1) \, \& \, B=(0.9045, 0, 0.8090)$, with $N=20$. We use the simplified notation $X \equiv \sigma^x_0$, $Y \equiv \sigma^y_0$, and $Z \equiv \sigma^z_0$. Note the range of the $y$ axis is very small. \label{fig:epsj_vs_m_extra_dE}] {\includegraphics[width=0.45\textwidth]{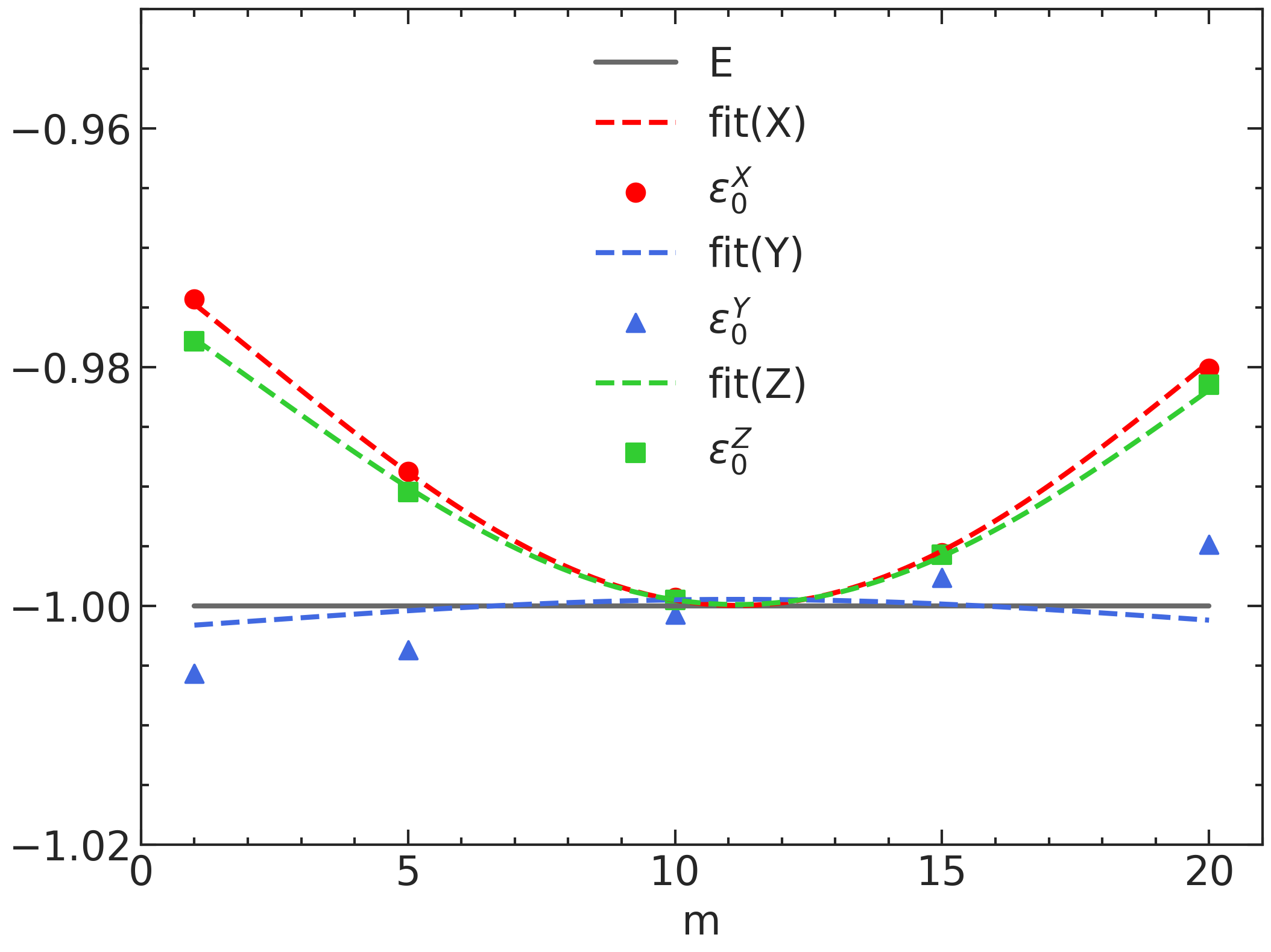}}
    \caption{We here give additional examples of plots of $\varepsilon_j^A(\theta_m)$ and $E(\theta_m)$ against the initial state, parametrized by $m$ (as described in subsection \ref{subsec:numerics_app}). In \ref{fig:epsj_vs_m_extra_2-body}, we show this plot for two-body observables for a specific model, whereas in \ref{fig:epsj_vs_m_extra_dE} we pick a model in which the average energy is constant and thus the energy standard deviation plays a bigger role. The coloured dashed lines are the fits to the data points given the function $\varepsilon_j = \gamma_j E + \eta_j \Delta E+ \chi_j$. Note that we are using analytical expressions for the energy average $E(\theta_m)$ and standard deviation $\Delta E(\theta_m)$, shown in appendix \ref{sec:analytical_formulas_E_dE}. Note also that we divided by $N$ the extensive quantities $\varepsilon_j, E, \Delta E$ to be able to compare them among different system sizes.} \label{fig:epsj_vs_m_extra}
\end{figure}

Finally, to complement the plot of $\varepsilon_j^A(\theta_m)$ and $E(\theta_m)$ against $m$ given in the main text, we here provide two additional examples. Fig.~\ref{fig:epsj_vs_m_extra_2-body} shows the plot for the same model previously considered, but this time for two-body observables. We note this is essentially the same behavior seen for one-body observables, and that this is true also for most other cases considered. On the other hand, the example shown in Fig.~\ref{fig:epsj_vs_m_extra_dE} corresponds to a model for which the energy is constant with respect to the initial state and therefore the dependence of $\varepsilon_j$ on the energy standard deviation is more significant.

\section{Error scaling in Trotterization} \label{sec:error_scaling_trotterization}
Given a Hamiltonian that is expressed as a sum of local terms, we want to justify the use of Trotterization to simulate the time evolution. 
We are considering Hamiltonians of the form \eqref{eqn_app:general_hamiltonian} with periodic boundary conditions, which indeed can be expressed as sums of local terms, i.e.,
\begin{equation}
    H = \sum_{i=0}^{N-1} h_i.
\end{equation}
Using the tools developed in \cite{childs_theory_2021_app} we can analyze the error of Trotterization on this class of Hamiltonians. Specifically, we know that the error $\epsilon$ scales as:
\begin{equation}
    \epsilon = \mathcal{O}\left(\frac{\alpha \,
    t_f^2}{r}\right),
\end{equation}
where $t_f$ is the simulation time, $r$ is the number of Trotter steps, and $\alpha$ is a Hamiltonian specific factor. In order to understand the scaling of the error as a function of $N$, we need to explicitly calculate $\alpha$. The definition of $\alpha$ is given as 
\begin{equation}
    \alpha := \sum_{p,q=1}^N ||[h_p, h_q]||.
\end{equation}
For the Hamiltonian we are concerned with, 
\begin{equation}
    ||[h_p, h_q]|| = \begin{cases}
0 &\text{$p=q$}\\
a &\text{$q=p\pm 1$} \\
0 &\text{otherwise}
\end{cases}
\end{equation}
where $a$ is just some scalar that depends on $J=(J_x,J_y,J_z)$ and $B=(B_x,B_y,B_z)$ but crucially does not depend on $N$. We see that the only contributions to $\alpha$ come from the commutator between terms in the Hamiltonian that have overlapping support. A simple counting argument shows that for $N$ sites, there are $2N$ combinations of overlapping terms. Therefore, $\alpha = 2aN$. This means that the error can be written as
\begin{equation}
    \epsilon = \mathcal{O}\left(\frac{2 a N \, t_f^2}{r}\right)
\end{equation}
We conclude that the Trotter error only grows linearly with the number of sites $N$.

\section{Analytical expressions for \texorpdfstring{$E(\theta_m)$}{E(θₘ)} and \texorpdfstring{$\Delta E^2(\theta_m)$}{Δ²E(θₘ)}} \label{sec:analytical_formulas_E_dE}
Here we provide the analytical expressions we have obtained for the average energy $E(\theta_m)$ and the energy variance $\Delta E^2(\theta_m)$ in terms of their dependence on $\theta_m$, which parametrizes the class of initial states that we are considering; see Eq.~\eqref{eqn_app:psi_theta_m}. For this family of initial states and for the class of Hamiltonians $H$ defined in Eq.~\eqref{eqn_app:general_hamiltonian}, we have that
\begin{equation}
    \begin{split}
     \frac{E(\theta_m)}{N} &\coloneqq \Tr(\rho_0(\theta_m) H) = \\
    &= -J^x \sin^2(\theta_m) -J^z \cos^2(\theta_m) \\ 
    &+ \frac{\llbracket \text{N is odd} \rrbracket}{N} [B^x \sin(\theta_m) + B^z \cos(\theta_m)]
    \end{split}
\end{equation}
and 
\begin{equation} \label{eqn_app:energy_variance_analytical_expression}
    \begin{split}
    \frac{\Delta E^2(\theta_m)}{N^2} &\coloneqq \Tr(\rho_0(\theta_m) H^2) - E^2(\theta_m) = \\
    &= \frac{1}{N} \sum_{\alpha=x,y,z} [(J^\alpha)^2 + (B^\alpha)^2] \\
    &+ \frac{2}{N} [(J^x)^2 + J^z J^y]\sin^2(\theta_m) \\
    &+ \frac{2}{N} [(J^z)^2 + J^x J^y]\cos^2(\theta_m) \\
    &- \frac{3}{N} [J^x \sin^2(\theta_m) + J^z \cos^2(\theta_m)]^2 \\
    &- \frac{1}{N} [B^x \sin(\theta_m) + B^z \cos(\theta_m)]^2 \\
    &- \frac{2 \llbracket \text{N is odd} \rrbracket}{N^2} [J^x B^x \sin(\theta_m) + J^z B^z \cos(\theta_m)]
    \end{split}
\end{equation}
where the Iverson bracket is defined as
\begin{equation*}
    \llbracket P \rrbracket \coloneqq 
    \begin{cases}
        1 &\; \text{if $P$ is true} \\
        0 &\; \text{otherwise}
    \end{cases}
\end{equation*}
and $N$ is the system size.

\section{Studying the coupling strength}

If a local subsystem of a non-integrable system thermalizes, then its reduced state is given by the Gibbs ensemble as long as the system has weak coupling, an exponential density of states, and large system size; see, e.g., Refs.~\cite{landau_statistical_2011_app, tolman_principles_1980_app, popescu_entanglement_2006_app}. Here we investigate whether the coupling is indeed weak enough that the Gibbs state can be assumed to be predictive, for the three non-integrable models considered. To do so, we will study three definitions of weak coupling found in the literature. \\

Consider the family of Hamiltonians given in Eq.~\eqref{eqn_app:general_hamiltonian}. Given a subsystem $S$, we can write the global Hamiltonian as $H = H_S \otimes \mathbb{I}_E + \mathbb{I}_S \otimes H_E + H_\textrm{int}$, where $H_S$ is the subsystem's Hamiltonian, $H_E$ is the effective environment's Hamiltonian and $H_\textrm{int}$ is the interaction Hamiltonian. For a one-site subsystem we have $H_S^{(1)} = \sum_\alpha B^\alpha \sigma^\alpha$ and $H_\textrm{int}^{(1)} = \sum_\alpha J^\alpha (\sigma^\alpha \otimes \sigma^\alpha \otimes \mathbb{I}_2 + \mathbb{I}_2 \otimes \sigma^\alpha \otimes \sigma^\alpha)$. For a two-site subsystem, instead we have $H_S^{(2)} = \sum_\alpha B^\alpha (\sigma^\alpha \otimes \mathbb{I}_2 + \mathbb{I}_2 \otimes \sigma^\alpha) + \sum_\alpha J^\alpha \sigma^\alpha \otimes \sigma^\alpha$ and $H_\textrm{int}^{(2)} = \sum_\alpha J^\alpha (\sigma^\alpha \otimes \sigma^\alpha \otimes \mathbb{I}_4 + \mathbb{I}_4 \otimes \sigma^\alpha \otimes \sigma^\alpha)$. As this discussion is only relevant for non-integrable systems, in the following we consider the three such models we have studied, namely, the Ising model with transverse and longitudinal fields, the XX model with transverse and longitudinal fields, and the XXZ model with transverse and longitudinal fields. The values of the parameters used can be found in Table~\ref{tab_app:models_params}. \\

To study the coupling strength, one first na\"ive option is to compare the operator norms of the subsystem and interaction Hamiltonians (e.g., as in Ref.~\cite{yunger_halpern_noncommuting_2020_app}). Analytically or numerically we compute $\| H_S^{(1)} \|$, $\| H_\textrm{int}^{(1)} \|$, $\| H_S^{(2)} \|$ and $\| H_\textrm{int}^{(2)} \|$; the results are shown in Table~\ref{tab_app:norms_gaps}. In particular, note that $\| H_S^{(1)} \| = \sqrt{\sum_\alpha (B^\alpha)^2}$ and $\| H_\textrm{int}^{(2)} \| = \sum_\alpha |J^\alpha|$. For the Ising model with transverse and longitudinal fields we can also compute exactly $\| H_\textrm{int}^{(1)} \| = 2$. We then calculate the ratios $\| H_\textrm{int}^{(1)} \| / \| H_S^{(1)} \|$ and $\| H_\textrm{int}^{(2)} \| / \| H_S^{(2)} \|$, which are shown in Table~\ref{tab_app:ratio_metrics}. Clearly the coupling cannot be considered weak according to this definition neither for one-site nor two-site subsystems. \\

A common definition based on perturbation theory is to compare the norms of the interaction Hamiltonians with the minimal difference $\Delta E_S$ between non-degenerate energy levels of the subsystem Hamiltonian~\cite{tasaki_quantum_1998_app, reimann_canonical_2010_app, riera_thermalization_2012_app, trushechkin_open_2022_app}. We find analytically $\Delta E_S^{(1)} = 2 \sqrt{\sum_\alpha (B^\alpha)^2}$ and compute $\Delta E_S^{(2)}$ numerically; the values for each non-integrable model are shown in Table~\ref{tab_app:norms_gaps}. We then compute the ratios $\| H_\textrm{int}^{(1)} \| / \Delta E_S^{(1)}$ and $\| H_\textrm{int}^{(2)} \| / \Delta E_S^{(2)}$, which are given in Table~\ref{tab_app:ratio_metrics}. Again, the coupling is clearly not weak also according to this definition. \\

\begin{table}[ht]
  \centering
  \begin{tabular}{lrrrrrr} 
    \toprule
    Model       & $\|H_S^{(1)}\|$ & $\|H_{\rm int}^{(1)}\|$ & $\|H_S^{(2)}\|$ & $\|H_{\rm int}^{(2)}\|$ & $\Delta E_S^{(1)}$ & $\Delta E_S^{(2)}$ \\
    \midrule
    Ising + LT  & $1.2135$      & $2.0000$             & $3.0693$      & $2.0000$             & $2.4270$          & $1.2340$          \\
    XX + LT     & $1.2135$      & $2.8284$             & $3.3033$      & $4.0000$             & $2.4270$          & $0.0693$          \\
    XXZ + LT    & $1.2135$      & $3.4780$             & $3.2943$      & $5.1778$             & $2.4270$          & $0.9392$          \\
    \bottomrule
  \end{tabular}
  \caption{Norms and minimal level spacings for the non‐integrable models.}
  \label{tab_app:norms_gaps}
\end{table}

\begin{table}[ht]
  \centering
  \begin{tabular}{lrrrr}
    \toprule
    Model       & $\displaystyle\frac{\|H_{\rm int}^{(1)}\|}{\|H_S^{(1)}\|}$
                & $\displaystyle\frac{\|H_{\rm int}^{(2)}\|}{\|H_S^{(2)}\|}$
                & $\displaystyle\frac{\|H_{\rm int}^{(1)}\|}{\Delta E_S^{(1)}}$
                & $\displaystyle\frac{\|H_{\rm int}^{(2)}\|}{\Delta E_S^{(2)}}$ \\
    \midrule
    Ising + LT  & $1.6481$      & $0.6516$             & $0.8241$          & $1.6207$          \\
    XX + LT     & $2.3308$      & $1.2109$             & $1.1654$          & $57.7140$         \\
    XXZ + LT    & $2.8661$      & $1.5717$             & $1.4330$          & $5.5130$          \\
    \bottomrule
  \end{tabular}
  \caption{Ratios of interaction strengths to subsystem norms and minimal energy gaps, for the non-integrable models.}
  \label{tab_app:ratio_metrics}
\end{table}

\begin{figure}
    \centering
    \includegraphics[width=\linewidth]{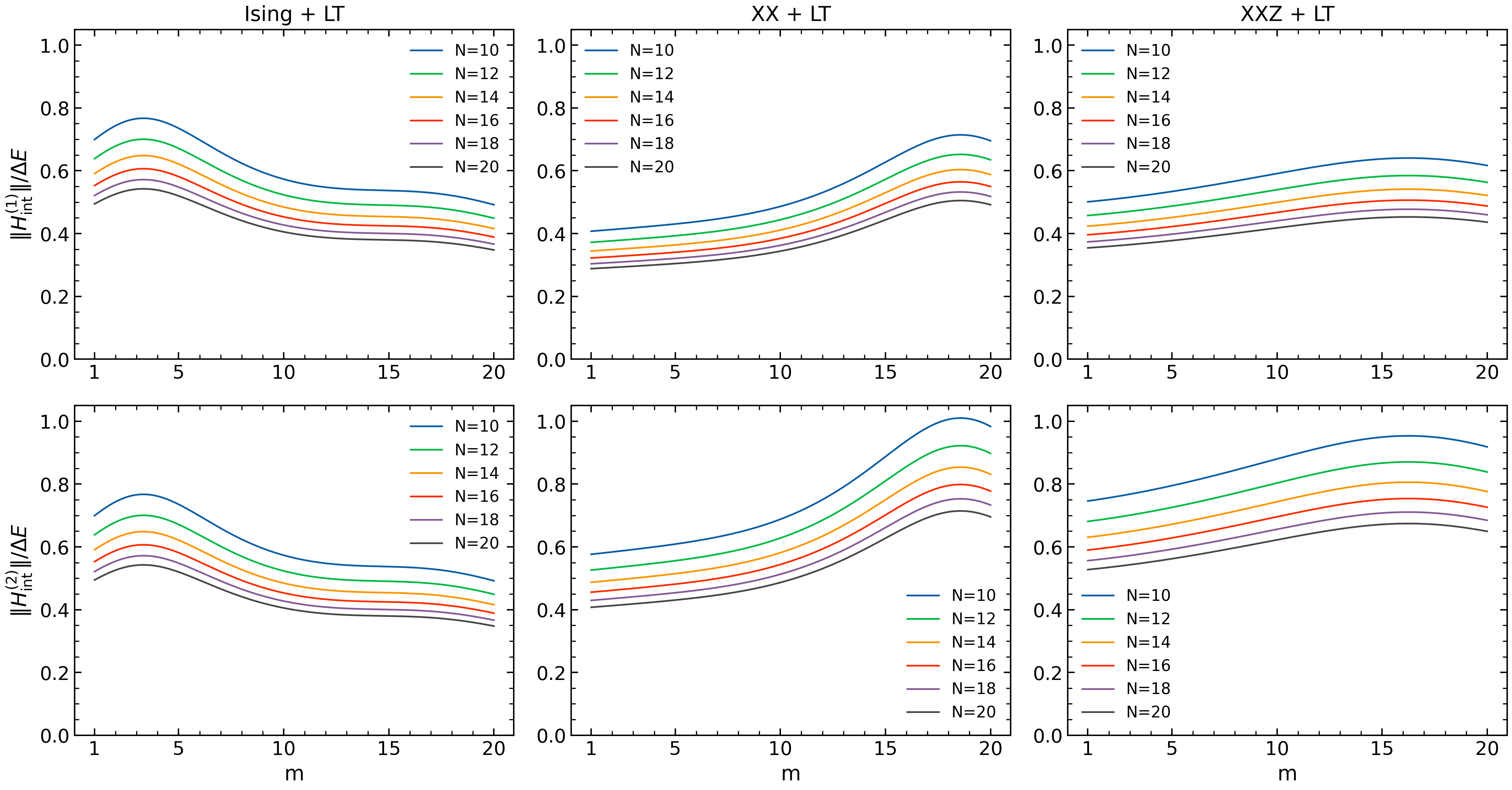}
    \caption{Ratios of interaction norms to the energy standard deviation plotted against the initial state parametrized by $\theta_m$, for all non-integrable models and system sizes considered.}
    \label{fig:ratios_Hint_dE_vs_theta_all_N_all_nonint_models}
\end{figure}

Finally, we consider a third definition, which is perhaps more physical \cite{reimann_canonical_2010_app, dong_quantum_2007_app, riera_thermalization_2012_app} and was suggested by Ref.~\cite{riera_thermalization_2012_app}, namely, 
\begin{equation}
    \| H_\textrm{int} \| \ll \Delta E,
\end{equation}
where $\Delta E$ is the standard deviation of the global Hamiltonian, i.e., the width of the microcanonical energy window. Having $\| H_\textrm{int} \| \approx \Delta E$ is sufficient to rule out weak coupling. For even $N$, the analytical expression for the standard deviation as a function of $\theta_m$ is
\begin{equation}
\begin{split}
    \Delta E(\theta_m) = \sqrt{N} \Bigg\{ &\sum_{\alpha=x,y,z} [(J^\alpha)^2 + (B^\alpha)^2]
    + 2 [(J^x)^2 + J^z J^y]\sin^2(\theta_m)
    + 2 [(J^z)^2 + J^x J^y]\cos^2(\theta_m) \\
    &- 3 [J^x \sin^2(\theta_m) + J^z \cos^2(\theta_m)]^2
    - [B^x \sin(\theta_m) + B^z \cos(\theta_m)]^2 \Bigg\}^{\displaystyle \frac{1}{2}}.
\end{split}
\end{equation}
In Fig.~\ref{fig:ratios_Hint_dE_vs_theta_all_N_all_nonint_models}, we plot the ratio of the operator norm of the interaction Hamiltonian to the standard deviation, namely, $\frac{\| H_\textrm{int}\|}{\Delta E}$, against the initial state parametrized by the angle $\theta_m$. The plots concern both one-site and two-site subsystems, and include data for all non-integrable models, system sizes and initial states considered. Even for the largest system size considered, i.e., $N=20$, the ratio is often greater than $0.5$ and in any case is always greater than $0.25$. Hence, in the parameter regime considered the coupling does not appear to be sufficiently weak to guarantee the predictivity of the local Gibbs ensemble. 

%

\end{document}